\pdfoutput=1

\documentclass[prd,twocolumn,twoside,preprintnumbers,superscriptaddress,nofootinbib]{revtex4}

\usepackage{amsmath}
\usepackage{graphicx}
\usepackage[hyperfootnotes=false]{hyperref}
\usepackage{xspace}
\usepackage{braket}

\newcommand{\be}{\begin{equation}}
\newcommand{\ee}{\end{equation}}

\newcommand{\eq}[1]{Eq.~\eqref{eq:#1}}
\newcommand{\eqs}[2]{Eqs.~\eqref{eq:#1} and \eqref{eq:#2}}
\newcommand{\eqss}[3]{Eqs.~\eqref{eq:#1}, \eqref{eq:#2}, and \eqref{eq:#3}}

\renewcommand{\sec}[1]{Sec.~\ref{sec:#1}}

\newcommand{\app}[1]{App.~\ref{app:#1}} 
\newcommand{\fig}[1]{Fig.~\ref{fig:#1}}
\newcommand{\Ref}[1]{Ref.~\cite{#1}}
\newcommand{\Refs}[1]{Refs.~\cite{#1}}
\newcommand{\Sec}[1]{Sec.~\ref{sec:#1}}

\newcommand{\pythia}{{\sc Pythia}\xspace}

\newcommand{\cuba}{{\sc CUBA}\xspace}

\newcommand{\ord}[1]{\mathcal{O}(#1)}

\newcommand{\df}{\mathrm{d}}

\newcommand{\sdt}{\!\cdot\!}
\newcommand{\tr}{\mathrm{tr}}
\newcommand{\nn}{\nonumber}
\newcommand{\cJ}{\mathcal{J}}
\newcommand{\cL}{{\mathcal L}}
\newcommand{\cusp}{\mathrm{cusp}}

\newcommand{\al}{\alpha}

\newcommand{\de}{\delta}
\newcommand{\De}{\Delta}
\newcommand{\eps}{\epsilon}
\newcommand{\ga}{\gamma}
\newcommand{\Ga}{\Gamma}
\newcommand{\si}{\sigma}

\newcommand{\zero}{{(0)}}
\newcommand{\one}{{(1)}}

\newcommand{\Mae}[3]{\bigl\langle#1\bigr\rvert#2\bigr\rvert#3\bigr\rangle}

\newcommand{\ORd}[1]{{\mathcal O}\Bigl(#1\Bigr)}

\newcommand{\bn}{\bar{n}}
\newcommand{\bnslash}{\bar{n}\!\!\!\slash}
\newcommand{\bnP}{\overline {\mathcal P}}
\newcommand{\cP}{{\mathcal P}}
\newcommand{\cB}{{\mathcal B}}

\newcommand{\lqcd}{\Lambda_\mathrm{QCD}}

\newcommand{\qL}{q^L}

\allowdisplaybreaks[4]

\begin{document}


\preprint{\vbox{\hbox{MIT--CTP 4476}}}

\title{Calculating Track Thrust with Track Functions}

\author{Hsi-Ming Chang}

\affiliation{Department of Physics, University of California at San Diego, La Jolla, CA 92093, USA}

\author{Massimiliano Procura}

\affiliation{Albert Einstein Center for Fundamental Physics, Institute for Theoretical Physics, University of Bern, CH-3012 Bern, Switzerland}

\author{Jesse Thaler}

\affiliation{Center for Theoretical Physics, Massachusetts Institute of Technology, Cambridge, MA 02139, USA}

\author{Wouter J.~Waalewijn}

\affiliation{Department of Physics, University of California at San Diego, La Jolla, CA 92093, USA}

\begin{abstract}

In $e^+e^-$ event shapes studies at LEP, two different measurements were sometimes performed:   a ``calorimetric'' measurement using both charged and neutral particles, and a ``track-based'' measurement using just charged particles.  Whereas calorimetric measurements are infrared and collinear safe and therefore calculable in perturbative QCD, track-based measurements necessarily depend on non-perturbative hadronization effects. On the other hand, track-based measurements typically have smaller experimental uncertainties. In this paper, we present the first calculation of the event shape \emph{track thrust} and compare to measurements performed at ALEPH and DELPHI.  This calculation is made possible through the recently developed formalism of track functions, which are non-perturbative objects describing how energetic partons fragment into charged hadrons.  By incorporating track functions into soft-collinear effective theory, we calculate the distribution for track thrust with next-to-leading logarithmic resummation.  Due to a partial cancellation between non-perturbative parameters, the distributions for calorimeter thrust and track thrust are remarkably similar, a feature also seen in LEP data.

\end{abstract}

\maketitle

\section{Introduction}

Detailed investigations of hadronic final states are crucial for understanding the dynamics of high-energy particle collisions.  Charged particles play a particularly important role in these investigations.  Whereas neutral particles can only be measured using calorimetry, charged particles can also be measured using tracking detectors, which allows for excellent momentum resolution and vertex identification.  At colliders like LEP, tracks were used to perform precision tests of quantum chromodynamics (QCD) through measurements of $e^+e^-$ event shapes and $N$-jet production rates~\cite{Buskulic:1992hq,Abreu:1996na} (see \Refs{Dasgupta:2003iq,Heister:2003aj,Abdallah:2003xz,Achard:2004sv,Abbiendi:2004qz} for reviews).  These LEP studies also tested hadronization models through measurements of charged hadron inclusive distributions.  Presently at the LHC, tracking information is used to improve jet measurements, to understand jet substructure, and to mitigate the effects of multiple ``pileup" collisions per single bunch crossing.

Despite the experimental advantages offered by tracks, most experimental and theoretical studies are aimed at infrared and collinear (IRC) safe observables, which include contributions from both neutral and charged particles.  In contrast, there are comparatively few theoretical tools available to understand and predict track-based observables.  While fragmentation functions (FFs) are useful for understanding the distribution of single charged particles, more general observables require non-perturbative information about charged particle correlations. For example, \Refs{Krohn:2012fg,Waalewijn:2012sv} showed how new non-perturbative functions are needed to calculate the energy-weighted charge of a jet.  Recently in \Ref{Chang:2013rca}, we introduced the formalism of \emph{track functions}, which enables QCD calculations to be performed on a broad class of track-based observables where (otherwise) IRC-safe observables are modified to include only charged particles. 

In this paper, we show how to use track functions to calculate track-based $e^+e^-$ event shapes in perturbative QCD.  The track function $T_i(x,\mu)$ is a non-perturbative object which describes how an energetic parton $i$ fragments to a collection of tracks carrying a fraction $x$ of the original parton energy \cite{Chang:2013rca}.  Like the FF and the jet charge distribution, the track function has a well-defined renormalization group (RG) evolution in $\mu$, such that one can measure $T_i(x,\mu)$ at one scale $\mu$ and use QCD perturbation theory to make predictions at another scale $\mu'$.  We will focus on the \emph{track thrust} event shape and compare our calculations to LEP measurements made by the ALEPH~\cite{Buskulic:1992hq} and DELPHI~\cite{Abreu:1996na} collaborations.

Our previous work in \Ref{Chang:2013rca} explained how to interface track functions with fixed-order calculations up to next-to-leading order (NLO). To get reliable predictions for track thrust, we need to include the effects of logarithmic resummation.  With the help of soft-collinear effective theory (SCET)~\cite{Bauer:2000ew,Bauer:2000yr,Bauer:2001ct,Bauer:2001yt}, we obtain results at next-to-leading logarithmic accuracy (NLL) including ${\cal O}(\alpha_s)$ fixed-order matching contributions, i.e.~up to NLL$'$ order.  This turns out to be sufficiently accurate to understand both the qualitative and quantitative behavior of the track thrust distribution.

We will show that ordinary (i.e.~calorimeter) thrust and track thrust are remarkably similar, with the leading differences encoded in a small number of non-perturbative parameters. Since an extraction of track functions from data has not yet been performed, we estimate these non-perturbative parameters using Monte Carlo event generators that have been tuned to LEP data (\pythia 8~\cite{Sjostrand:2006za,Sjostrand:2007gs} in this study).  We find cancellations between the non-perturbative parameters, such that the predicted distributions for calorimeter thrust and track thrust are nearly identical, a feature also seen in LEP data.  This behavior could have been anticipated based on the observation in \Ref{Chang:2013rca} that hadronization effects are strongly correlated between the numerator and denominator of dimensionless track-based ratios.  We can now put this qualitative observation on a firmer quantitative footing.

An interesting theoretical feature of our calculation is that hadronization effects enter directly into the track thrust resummation.  In particular, non-perturbative track parameters appear in the anomalous dimensions of the (track-based) jet and soft functions, two important objects in the factorization theorem for the track thrust distribution.  As a nice consistency check of our formalism, we find that the hard, jet, and soft anomalous dimensions still cancel, despite the appearance of these parameters. We also show how to incorporate the leading non-perturbative power correction in the track thrust distribution.

This paper is structured as follows. \sec{summary} contains a summary of our results and the most significant plots, including a comparison to LEP data. The underlying technical details are discussed in the rest of the paper. We review our track function formalism in \sec{tracks} and calculate track thrust at $\ord{\alpha_s}$ in \sec{NLO}.  In \sec{resum} we present the factorization theorem for track thrust as well as the ingredients needed for a resummation up to NLL$'$ order in SCET, with details on the RG evolution given in the appendices. A simple expression for track thrust at NLL order is derived in \sec{NLL}, which allows us to better understand the similarity between calorimeter and track thrust.  Our final numerical results are presented in \sec{results}.  We conclude  in \sec{discussion} with a discussion of possible generalizations of our results to other track-based observables.

\section{Summary of Results}
\label{sec:summary}

To begin, we define the two main event shapes used in our study:  calorimeter thrust $\tau$ and track thrust $\overline{\tau}$.  The classic event shape thrust~\cite{Farhi:1977sg} is defined as
\be
T = \max_{\hat t}\, \frac{\sum_{i} |\hat t \sdt {\vec p}_i|}{\sum_{i} |{\vec p}_i|},
\ee
where the sum runs over all final-state hadrons with momenta ${\vec p}_i$, and the unit vector $\hat{t}$ defines the thrust axis.  It is more convenient to work with
\be
\label{eq:thrustdef}
\tau \equiv 1- T = \min_{\hat t} \frac{\sum_{i}  \left( |{\vec p}_i| - |\hat t \sdt {\vec p}_i| \right) }{\sum_{i} |{\vec p}_i|}\, , 
\ee
which we will refer to as ``thrust'' from now on.
Since this is measured using all final-state hadrons (charged plus neutral), we call $\tau$ \emph{calorimeter thrust}. \textit{Track thrust} $\bar{\tau}$ is defined analogously to \eq{thrustdef}, except that the sum over $i$ is restricted to charged particles in both the numerator and the denominator.  In this paper, a bar will always indicate a track-based quantity.

\begin{figure}
\includegraphics[height=45ex]{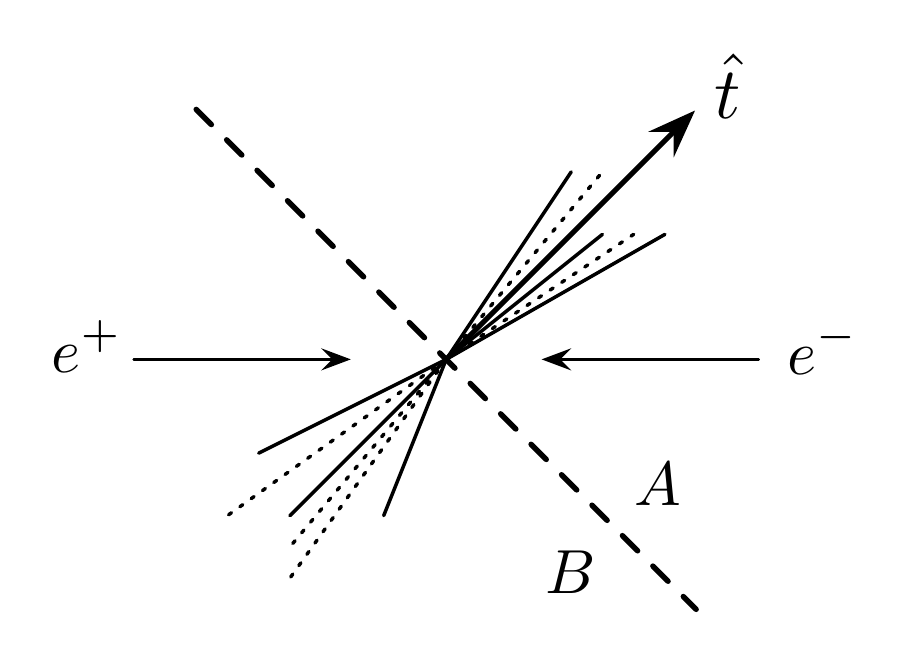} \\
\caption{Illustration of the track thrust measurement in an $e^+ e^-$ event with jets initiated by a $q \bar q$ pair.  Solid lines indicate charged particles and dashed lines indicate neutral particles.  For track thrust, the thrust axis $\hat{t}$ is determined by the charged particles alone.  The event is divided into hemispheres $A$ and $B$ by a plane perpendicular to the thrust axis.}
\label{fig:trackthr}
\end{figure}

For the later discussion of the factorization theorem for track thrust in \Sec{resum}, it will be convenient to rewrite thrust in terms of contributions from hemispheres $A$ and $B$, separated by a plane perpendicular to the thrust axis.  The relevant kinematics are illustrated in \fig{trackthr}.
Fixing two light-cone vectors $n^\mu$ and $\bn^\mu$ such that $n \cdot \bn = 2$, the light-cone components of any four-vector $w^\mu$ are given by $w^+ = n\sdt w$, $w^- = \bn\sdt w$, and $w_\perp^\mu$, such that
\begin{equation} \label{eq:LC}
  w^\mu = w^+ \frac{\bn^\mu}{2} + w^- \frac{n^\mu}{2} + w_\perp^\mu
\,.
\end{equation}
Choosing $n^\mu=(1,0,0,1)$ and $\bn^\mu= (1,0,0,-1)$ with the 3-axis aligned along $\hat t$, we can rewrite \eq{thrustdef} for tracks as
\begin{align}
\label{eq:bartau}
  \bar \tau = \frac{2\,(\bar k_A^+ + \bar k_B^-)}{(x_A + x_B)\,Q}
\,.
\end{align}
Here, $Q$ is the $e^+ e^-$ center-of-mass energy, $x_{A,B}$ are the energy fractions of charged particles in the respective hemispheres, and $\bar k_A^+ ={\bar k}_A^0-{\bar k}_A^3$ and $\bar k_B^-={\bar k}_B^0+{\bar k}_B^3$ are the small light-cone momentum components of all the charged particles in hemisphere $A$ and $B$, respectively. In this paper, we ignore the subtleties of hadron masses and measurement schemes, which will affect power corrections (see~\Refs{Salam:2001bd,Mateu:2012nk}).

\begin{figure}
\includegraphics[height=40ex]{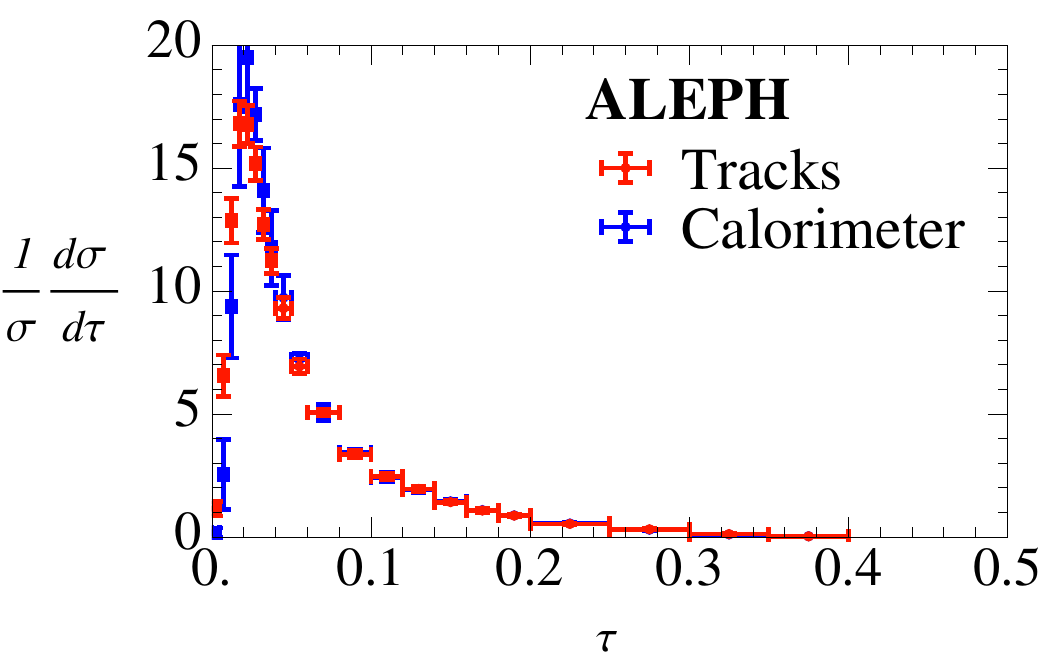} \\
\includegraphics[height=40ex]{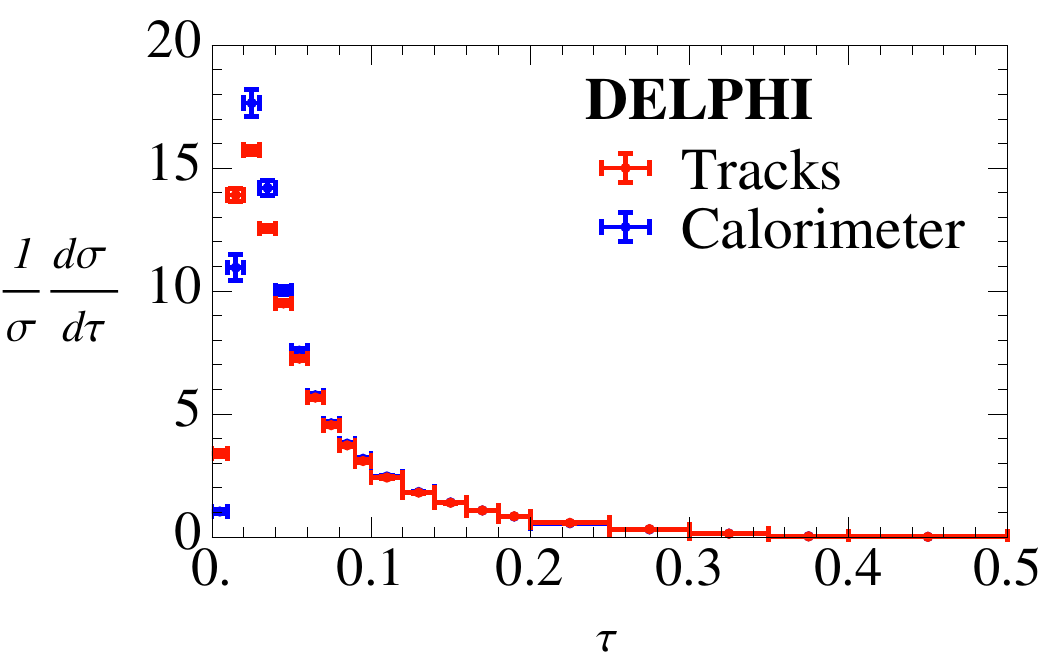}
\caption{ALEPH (top) and DELPHI (bottom) measurements of calorimeter and track thrust.  Error bars correspond to the statistical and systematic uncertainties added in quadrature.  The experimental uncertainties associated with the track-based measurements are noticeably smaller.}
\label{fig:rawdata}
\end{figure}

At LEP, differential cross sections for calorimeter thrust  $\tau$ and track thrust  $\bar \tau$ were measured at both ALEPH \cite{Buskulic:1992hq} and DELPHI \cite{Abreu:1996na} on the $Z$ pole ($Q = 91\; \text{GeV}$).  (To our knowledge, these are the only two experiments with public data on track thrust.) In both experiments, measurements were unfolded to the hadron level (including both charged and neutral hadrons for  $\tau$, and only charged hadrons for $\bar \tau$).  
The ALEPH and DELPHI normalized distributions are shown in \fig{rawdata}, where we note a remarkable similarity between the calorimetric and track-based measurements. Indeed, for all bins outside of the peak region, the distributions are compatible within error bars, and a key goal of this paper is to gain an analytic understanding for why the $\tau$ and $\bar \tau$ distributions are so similar. Note also that the experimental uncertainty is significantly smaller for the thrust measurements made using tracks.

\begin{figure}
\includegraphics[height=40ex]{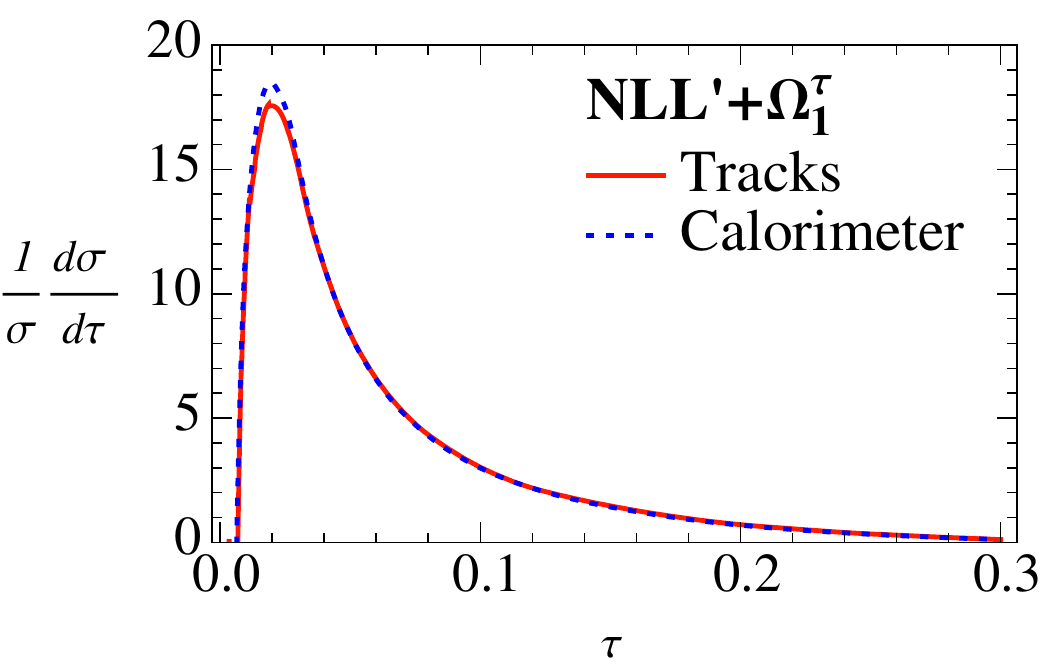} \\
\includegraphics[height=40ex]{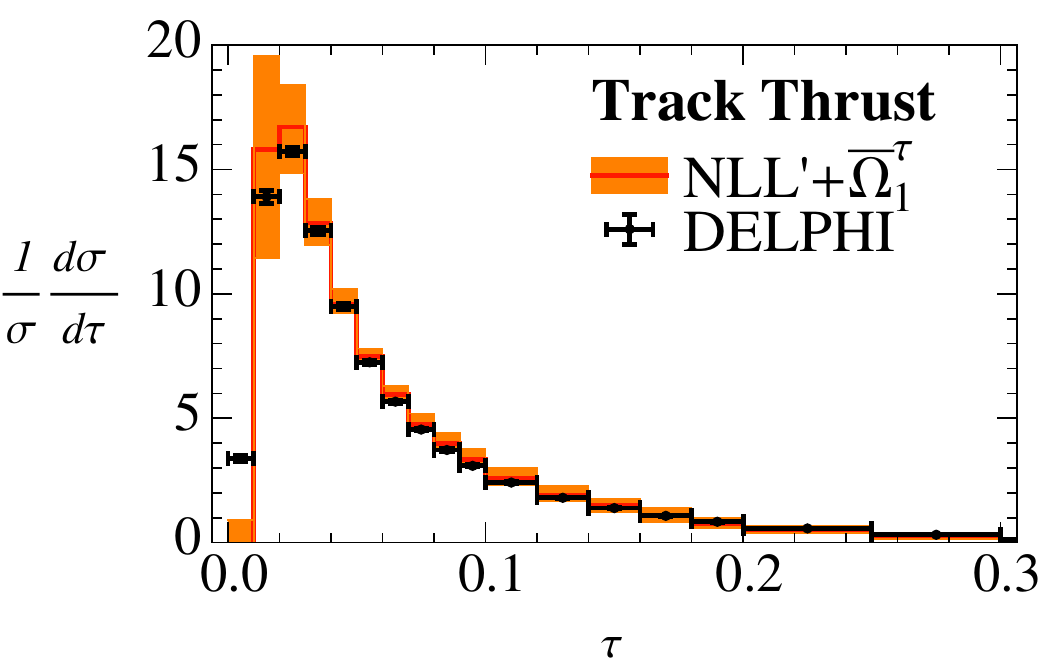}
\caption{Top: NLL$'$ distributions for calorimeter and track thrust including the leading non-perturbative correction $\Omega_1^\tau$. Next-to-leading logarithmic resummation is included together with ${\cal O}(\alpha_s)$ fixed-order matching contributions.  The NLL$'$ calculation exhibits the same qualitative similarity between calorimeter and track thrust as seen in LEP data.  Bottom: comparing our analytic results to the DELPHI measurement.  There is good quantitative agreement in the tail region where our NLL$'$ calculation is most accurate.  The theoretical uncertainties are from scale variation alone, and do not include the (correlated) uncertainties in $\alpha_s$ or  $\Omega_1^\tau$, nor uncertainties in our track function extraction.}
\label{fig:summary}
\end{figure}

In \fig{summary}, we show the main result of the paper: the resummed NLL$'$ distributions for calorimeter and track thrust. The latter was obtained using track functions extracted from \pythia 8, which itself was tuned to LEP data. The effects of the leading non-perturbative power correction are included through the parameters $\Omega^\tau_1$ and $\bar{\Omega}^\tau_1$, which are different for calorimeter and track thrust.  Interestingly, the NLL$'$ distributions exhibit the qualitative similarity seen in data between calorimeter thrust and track thrust.  We also see excellent quantitative agreement between our result and DELPHI measurements in the peak and tail regions. To the left of the peak there are deviations due to important non-perturbative corrections and in the far-tail region our calculation is missing (known) higher-order perturbative effects. 

We now briefly discuss why the $\tau$ and $\bar \tau$ distributions are so similar, referring the reader to \sec{NLL} for further details.  In \eq{bartau}, the numerator is dominated by soft gluon emissions which broaden the hemisphere jets, whereas the denominator is mainly affected by fragmentation of the energetic quark and antiquark emerging from the underlying scattering process. These effects are thus controlled by different track functions (gluon vs.~quark) but nearly cancel each other out due to the specific form of the (\pythia-based) track functions.

This cancellation is best understood by studying the resummed form of cumulative distributions
\be
\label{eq:cumulative}
\Sigma(\tau^c) \equiv  \int_0^{\tau^c}\!\!\! \df \tau\, \frac{\df \si}{\df \tau}, \qquad
\bar \Sigma(\bar \tau^c) \equiv  \int_0^{\bar \tau^c}\!\!\! \df \bar \tau\, \frac{\df \si}{\df \bar \tau}.
\ee
As we show in \sec{NLL}, at NLL the difference between the cumulative distributions (for $\tau^c<1/3$) is almost entirely captured by
\be
\label{eq:approxNLLcumulative}
\bar \Sigma(\bar \tau^c) \simeq \Sigma(\bar \tau^c) \times (3\bar \tau^c)^{\De},
\ee
where the exponent $\Delta$ redistributes the cross section between the peak and tail regions. In terms of the strong coupling constant $\alpha_s$ and the quark color-factor $C_F = 4/3$, the explicit form of $\Delta$ is
\be
\Delta = \frac{2 \alpha_s C_F}{ \pi} \left(g_1^L - \qL \right),
\ee
which depends on just two non-perturbative parameters:  a logarithmic moment of a single gluon track function $g_1^L$ and a logarithmic moment of two quark track functions $\qL$.  The similarity between the $\tau$ and $\bar \tau$ distributions can thus be traced to a cancellation between $g_1^L$ and $\qL$ such that $|\Delta| \simeq 0.004$ (see \eq{logmoments}).  

There are additional effects at NLL$'$ from the fixed-order matching which yield further (small) differences between $\tau$ and $\bar \tau$ which are compatible with the ALEPH and DELPHI measurements.  The non-perturbative power corrections $\Omega_1^\tau$ and $\bar \Omega_1^\tau$ lead to a respective shift of the $\tau$ and $\bar{\tau}$ distributions by a very similar amount, but increase the difference in the peak region.  Overall, though, the similarity between calorimeter and track thrust is well-described by the NLL distribution, and we expect similar cancellations to occur for a variety of (dimensionless) track-based observables.

\section{Review of Track Function Formalism}
\label{sec:tracks}

A rigorous QCD description of track-based observables involves \emph{track functions} $T_i(x,\mu)$~\cite{Chang:2013rca} as key ingredients. A parton (quark or gluon) with flavor index $i$ and four-momentum $p_i^\mu$ hadronizes into charged particles (tracks) with total four-momentum $\overline{p}_i^\mu \equiv x p_i^\mu+\ord{\Lambda_\text{QCD}}$. The track function is the distribution in the energy fraction $x$ of all tracks (irrespective of their multiplicity or individual properties), and it is normalized as
\begin{align} \label{eq:T_norm}
 \int_0^1\! \df x\ T_i(x,\mu) = 1
\,.\end{align}
We will often refer to $x$ as the track fraction.

In the context of factorization theorems, track functions can be used for track-based observables where partons in the underlying process are well-separated, i.e.~where their typical pairwise invariant masses are larger than $\lqcd$.  In this limit, each parton has its own independent track function, with correlations captured by power corrections (to be discussed more in \Sec{powercorrections}).  The track functions then encode process-independent non-perturbative information about the hadronization.  Like a FF or a parton distribution function (PDF), $T_i(x,\mu)$ absorbs infrared (IR) divergences in partonic calculations. It has a well-defined dependence on the RG scale $\mu$ through an evolution equation which is closely reminiscent of the jet charge distribution~\cite{Waalewijn:2012sv}. 

QCD calculations of track-based observables require the determination of matching contributions from partonic cross sections. First recall that the cross section for an IRC safe observable $e$ measured using partons has the form
\begin{align}
\frac{\df \sigma}{\df e} = \sum_N \int\! \df \Pi_N\, \frac{ \df \si_N}{\df \Pi_N}\, \de[e - \hat{e}(\{p^\mu_i\})]
\,,\end{align}
where we drop possible convolutions with PDFs to keep the notation simple.  Here, $\Pi_N$ denotes the $N$-body phase space, $\df \si_N/\df \Pi_N$ is the corresponding partonic cross section, and $\hat{e}(\{p_i\})$ implements the measurement on the partonic four-momenta $p^\mu_i$.  Since $e$ is IRC safe, a cancellation of final state IR divergences between real and virtual diagrams is guaranteed by the KLN theorem~\cite{Kinoshita:1962ur,Lee:1964is}.  

For the same observable measured using only tracks, we can write the cross section in the form
\begin{equation} \label{eq:matching}
\frac{\df \sigma}{\df \overline{e}}= \sum_N \int\! \df \Pi_N\, \frac{\df \bar \si_N}{\df \Pi_N} \int\! \prod_{i=1}^N \df x_i\, T_i(x_i)\, \de[\bar e - \hat{e}(\{x_i p^\mu_i\})].
\end{equation}
Here, the partonic cross section $\bar \si_N$ should be thought of as a finite matching coefficient where the IR divergences in $\si_N$ have been removed using some scheme.  These IR (collinear) divergences are absorbed by the track function $T_i(x_i)$ (which is similarly scheme-dependent).  The universality of collinear divergences in QCD~\cite{Berends:1987me,Mangano:1990by,Kosower:1999xi} guarantees the feasibility of this matching to all orders in $\alpha_s$.  In \Ref{Chang:2013rca} we explicitly showed the cancellation of IR-divergent terms in the partonic cross section $e^+ e^- \to q \bar{q} g$, which enters the NLO distribution for the energy fraction of charged particles in $e^+ e^-$ collisions. 

The (bare) track function is defined in QCD in a fashion analogous to the unpolarized FF (cf.~\cite{Collins:1981uk,Collins:1981uw}). Expressed in terms of light-cone components (see \eq{LC}), the quark track function is
\begin{align} \label{eq:defTqQCD} 
 T_{q}(x)&=
\int\! \df y^+\,\df^2 y_\perp\;e^{\,i k^- \, y^+/2}  \, \frac{1}{2 N_c}\, \sum_{C, N} \de\Big(x - \frac{p_C^-}{k^-}\Big)
\nn \\ & \, \, \times \tr\Big[ \frac{\bnslash}{2}\,
  \langle 0 | \psi (y^+, 0, y_\perp )| C N
  \rangle \langle C N | \overline{\psi}(0) |0 \rangle \Big]
\,,\end{align}
where $\psi$ is the quark field, $C$ ($N$) denote charged (neutral) hadrons, and $p_C^-$ is the large momentum component of all charged particles. As for the FF, gauge invariance requires the addition of eikonal Wilson lines. The factor $1/(2 N_c)$ in \eq{defTqQCD} comes from averaging over the color and spin of the hadronizing quark. The gluon track function is defined analogously. In $d$ space-time dimensions,
\begin{align} \label{eq:defTgQCD} 
T_g(x)& = - \frac{1}{(d-2) (N_c^2-1) k^-} 
 \\ & \, \, \times \int\! \df y^+\,\df^2 y_\perp\;e^{\,i k^- \, y^+/2}  \, \sum_{C, N} \de\Big(x - \frac{p_C^-}{k^-}\Big)
\nn \\ &\,\,\times \,\bar{n}^\mu \bar{n}^\nu
  \langle 0 |G^a_{\mu \lambda}(y^+, 0, y_\perp )| C N
  \rangle \langle C N |  G^{\lambda,a}_\nu(0) |0 \rangle,
\nn\end{align}
where $G_{\mu \nu} = \sum_a G_{\mu \nu}^a\, T^a$ is the QCD field-strength tensor and an average over colors and the $(d-2)$ polarizations of the gluon is performed.

For the sake of completeness, we also give SCET expressions for the quark and gluon track functions, given in a form which is invariant under non-singular gauge transformations. In terms of the SCET $n$-collinear quark $\chi_n(y)$ and gluon $\cB_{n\perp}^\mu(y)$ fields, we obtain
\begin{align} \label{eq:TqSCETdef}
  T_q(x) & = 
  2(2 \pi)^3\, \frac{1}{2 N_c}\,\sum_{C, N}\,\de\Big(x - \frac{ p_C^-}{k^-}\Big)
  \nn \\ & \,\, \times \tr
  \Big [\frac{\bnslash}{2}  \Mae{0}{[\de(k^- - \bnP) \,\de^2(\cP_\perp) \chi_n(0)]} {C N}
   \nn \\
  &\,\, \times
  \Mae{C N}{\bar \chi_n(0)}{0} \Big ], 
\end{align}
and
\begin{align} \label{eq:TgSCETdef}
  T_g(x) & = 
  - 2(2 \pi)^3\, \frac{k^-}{(d-2) (N_c^2-1)}\sum_{C, N}\,   \de\Big(x - \frac{p_C^-}{k^-}\Big)
  \nn \\
  &\,\, \times \Mae{0}{[\de(k^- - \bnP) \,\de^2(\cP_\perp) \cB_{n\perp}^{\mu, a}(0)]} {C N}
   \nn \\
  &\,\, \times
  \Mae{C N}{\cB_{n\perp,\mu}^a(0)}{0}
   \,,
\end{align}
where the momentum operators $\bnP = \bn \sdt \cP$ ($\cP_{\perp}^\mu$) return the sum of the minus (perpendicular) \emph{label} momentum components of all collinear fields on which they act. For the definition of the SCET fields, we refer the reader to e.g.~\Ref{Jain:2011xz}. 

\begin{figure}
\includegraphics[width=0.4\textwidth]{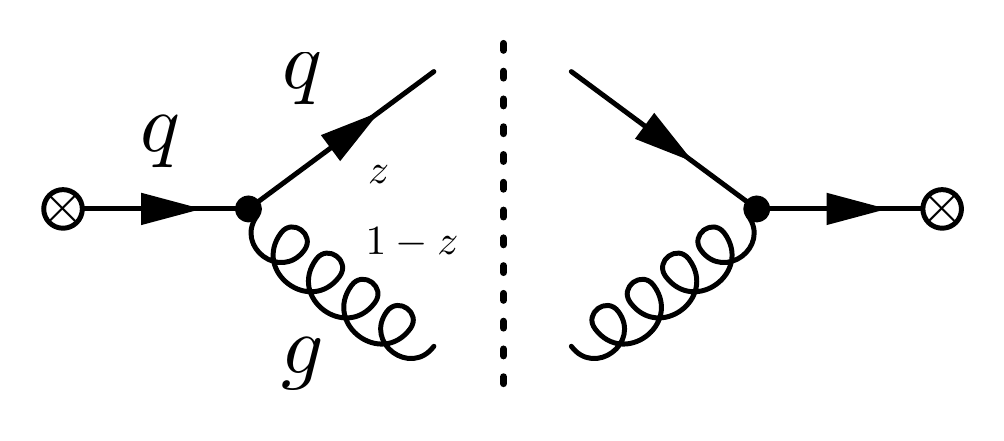} \\
\includegraphics[width=0.4\textwidth]{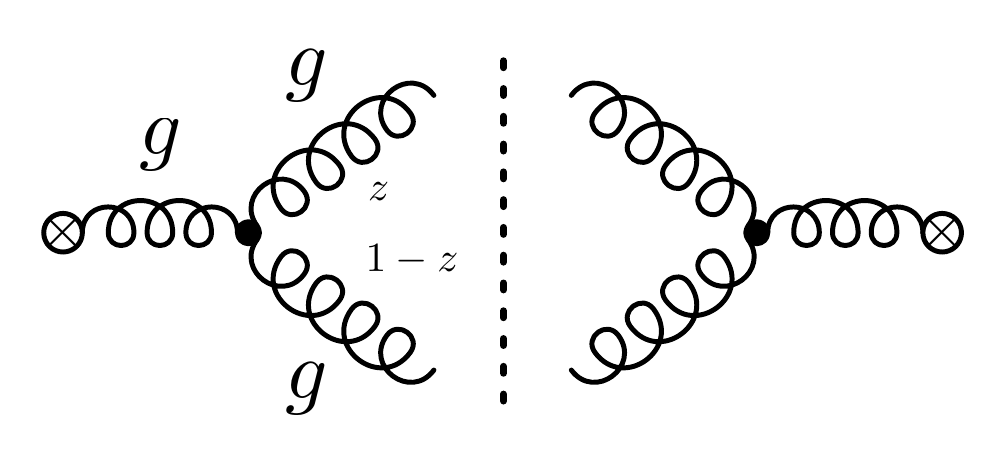}\\
\includegraphics[width=0.4\textwidth]{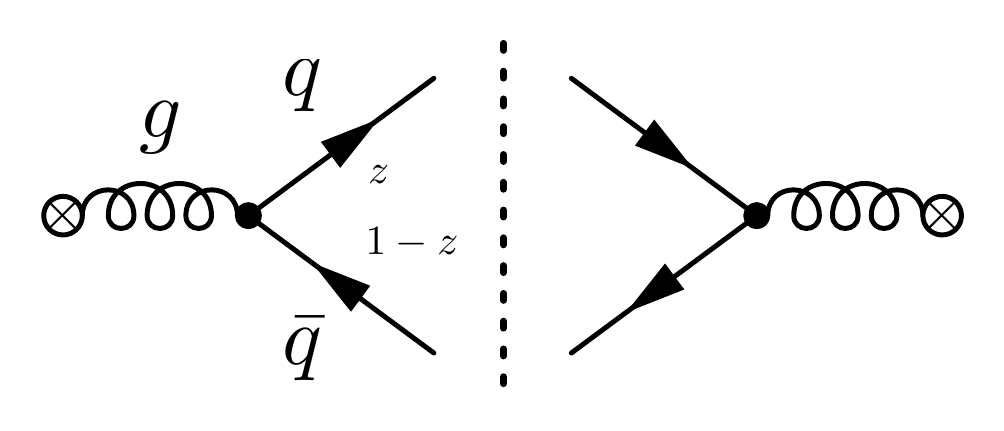}
\caption{Perturbative QCD calculation of the quark (top) and gluon (middle and bottom) track functions at NLO from \eqs{defTqQCD}{defTgQCD}  with partonic intermediate states. The NLO track function gets contributions from both branches of the collinear splitting. We do not display virtual diagrams, which vanish in pure dimensional regularization, or diagrams corresponding to Wilson line emissions.}
\label{fig:NLOsplitting}
\end{figure}

Although the track function is a non-perturbative object, some of its properties can be calculated in perturbation theory. In particular, the RG evolution of the track function follows from its ultraviolet (UV) divergences, as we show below. A partonic calculation of the track function is also necessary for extracting the matching coefficient $\bar \si_N$ in \eq{matching}, by using that this equation holds at both the hadronic and partonic level.

At NLO, we can relate the bare track function $T_{i,\text{bare}}^\one(x)$ to the tree-level track functions $T^\zero_j(x_1)$ and $T^\zero_k(x_2)$ via a collinear splitting $i \to jk$. 
As indicated in \fig{NLOsplitting}, this splitting is controlled by the timelike Altarelli-Parisi splitting functions $P_{i \to jk}(x)$~\cite{Altarelli:1977zs}.  In pure dimensional regularization with $d = 4 - 2 \epsilon$,
\begin{align} \label{eq:Toneloop}
T_{i,\text{bare}}^\one(x) & = \frac{1}{2} \sum_{j,k} \int_0^1\! \df z \, \Big[\frac{\al_s(\mu)}{2\pi} \Big(\frac{1}{\eps_\text{UV}} - \frac{1}{\eps_\text{IR}} \Big) P_{i \to jk}(z)\Big] 
\nn \\ & \quad \times 
\int \! \df x_1 \, \df x_2 \,  T^\zero_j(x_1)\, T^\zero_k(x_2) 
\nn \\ & \quad \times 
\de\big[x - z x_1 - (1-z) x_2\big]
\,. \end{align}
If $j=k$, the factor 1/2 is needed for identical particles, whereas if $j \neq k$ this factor gets cancelled by permutations of the two indices.  In contrast to the FF or PDF, the NLO track function gets contributions from both branches of the splitting.  

Renormalizing the UV divergences in the $\overline{\text{MS}}$-scheme leads to the following evolution equation for the track function,
\begin{align}\label{eq:T_RGE}
  \mu \frac{\df}{\df \mu}\, T_i(x, \mu) &= \frac{1}{2} \sum_{j,k} \int_0^1\! \df z\, \df x_1\, \df x_2\, \frac{\al_s(\mu)}{\pi} P_{i \to j k}(z)
  \nn \\ & \quad \times 
  T_j(x_1,\mu) T_k(x_2,\mu)
    \nn \\ & \quad \times 
  \de[x \!-\! z x_1 \!-\! (1\!-\!z) x_2]. 
\end{align}
By solving this, $T_i(x, \mu)$ can be extracted at one scale and RG evolved to another scale, and the evolution preserves the normalization in \eq{T_norm}. We note that the number of convolutions in the track function RG equation (RGE) grows accordingly to the perturbative order due to multiple branchings, so it becomes numerically more involved to solve this RGE at higher orders.  At leading logarithmic (LL) accuracy, the RG evolution in \eq{T_RGE} is equivalent to a parton shower~\cite{Waalewijn:2012sv}, and is in excellent agreement with the parton shower evolution in \pythia~\cite{Chang:2013rca}.

Throughout this paper, we determine the track functions used in our analytic formulae using the method of \Ref{Chang:2013rca}.  That is, we generate pure quark and gluon jet samples with \pythia 8.150~\cite{Sjostrand:2006za,Sjostrand:2007gs}, measure the normalized distribution for the track fraction $x$ within those jets, and extract the track functions by numerically inverting the analytic expression for the same quantity at either LO or NLO.  In all of the plots shown here, we use NLO track functions.  We emphasize that the use of \pythia\ is not fundamental, and one could imagine extracting the same information from $e^+ e^-$ data.  That said, since \pythia\ is tuned to LEP data, we expect these track functions to be realistic, but we have not attempted to assign uncertainties to the track functions.

One important point is the choice of $\alpha_s$.  Since we are working at NLL$'$ order in the $\overline{\text{MS}}$ scheme, it would be natural to take the value from \Ref{Abbate:2010xh} of $\alpha_s(M_Z) = 0.1203 \pm 0.0079$.  However, we have extracted the track functions from \pythia~8 whose default value is $\alpha_s(M_Z) = 0.1383$ for the final state parton shower, leading to a formal mismatch between our perturbative and non-perturbative objects.  Given the large uncertainties at NLL$'$, we will make an (imperfect) compromise, and extract the NLO track functions from \pythia\ using \pythia's value of $\alpha_s$, but then use
\be
\label{eq:alphasvalue}
\alpha_s(M_Z) = 0.125,
\ee
for all subsequent calculations.  This choice, along with the leading power correction in \Sec{powercorrections}, gives a good description of the LEP calorimeter thrust data.  As emphasized in \Ref{Abbate:2010xh}, there are strong correlations between the value of $\alpha_s$ and the leading power correction $\Omega_1^\tau$, so there are many different choices which would give comparable results; for example the \pythia\ value $\alpha_s(M_Z) = 0.1383$ matches the LEP calorimeter thrust distributions quite well with $\Omega_1^\tau = 0$.  A proper treatment of the correlations between these parameters is beyond the scope of this paper, so we will not show the uncertainties associated with $\alpha_s(M_Z)$ or $\Omega_1^\tau$.

\section{Fixed Order Analysis of Track Thrust}
\label{sec:NLO}

The leading non-trivial process for thrust at the partonic level is $e^+ e^- \to q \bar{q} g$, which appears at $\ord{\alpha_s}$ in a fixed-order expansion.  Given an $e^+ e^-$ collision at a center-of-mass energy $Q$, the kinematics of this process are determined by the partonic energy fractions $y_i=2E_i/Q$ carried by the quark and antiquark, with the gluon energy fraction given by $y_3 = 2 - y_1 - y_2$.  From this information, one can readily find the three-momenta of the partons $\vec{p}_1$, $\vec{p}_2$, and $\vec{p}_3$ and determine calorimeter thrust from \eq{thrustdef}.  For three partons, finding the thrust axis is straightforward, and thrust takes a reasonably simple form
\begin{equation}
\label{eq:threpartonthrust}
\tau = 1 - \frac{\max_{i = 1,2,3} |\vec{p}_{\rm CM} - 2 \vec{p}_i|} {\sum_i |\vec{p}_i| },
\end{equation}
where we have defined
\begin{equation}
\vec{p}_{\rm CM} \equiv \vec{{p}}_1 + \vec{{p}}_2 + \vec{{p}}_3.
\end{equation}

To obtain the charged track three-momenta, one simply rescales the parton momenta by the track fraction $x_i$,
\begin{equation}
\vec{\bar{p}}_i = x_i \vec{p}_i.
\end{equation}
Track thrust can then be calculated from \eq{threpartonthrust} with all $\vec{p}$ replaced by $\vec{\bar{p}}$.  Note that in the $e^+ e^-$ rest frame, 
$|\vec{p}_{\rm CM}| = 0$, but $|\vec{\bar{p}}_{\rm CM}|$ is typically non-zero.

The calculation of the track thrust distribution at $\ord{\alpha_s}$ is very similar to the one performed in \Ref{Chang:2013rca} for the total charged particle energy fraction.  Weighting each parton by the corresponding track function, we find
\begin{align} \label{eq:NLO}
  \!\!\frac{\df \si}{\df \bar \tau} &= \int_0^1\! \df y_1 \df y_2\, \frac{\df \bar \si(\mu)}{\df y_1 \df y_2} 
  \int_0^1 \! \df x_1 \df x_2  \df x_3 \,T_q(x_1,\mu) T_q(x_2,\mu) 
  \nn \\ & \quad \times
  T_g(x_3,\mu)\, \de[\bar \tau - \bar \tau(y_1,y_2,x_1,x_2,x_3)]
\,. \end{align}
where the measurement function $\bar \tau(y_1,y_2,x_1,x_2,x_3)$ implements \eq{threpartonthrust}.  Note that $T_q = T_{\bar q}$, by charge conjugation.  The relevant doubly differential partonic cross section is given in \Ref{Chang:2013rca} in the $\overline{\text{MS}}$ scheme.  Ignoring the singularities at $y_1 = 1$ and $y_2 = 1$ (which only contribute to a delta function at $\bar \tau = 0$),
\be
 \frac{\df \bar \si(\mu)}{\df y_1 \df y_2} =  \si_0 \, \frac{\al_s(\mu) C_F}{2\pi}
   \frac{\theta(y_1+y_2-1)(y_1^2 + y_2^2)}{(1-y_1) (1-y_2)} + \ldots.
\ee
Here, $\si_0$ is the total Born cross section
\begin{align}  \label{eq:Bornxsec}
\si_0 &= \frac{4\pi \al^2 N_c}{3Q^2} 
\\
&\quad \times \biggl( Q_q^2 \!+\! \frac{(v_q^2 \!+\! a_q^2) (v_\ell^2 \!+\! a_\ell^2) \!-\! 2 Q_q v_q v_\ell (1\!-\!M_Z^2/Q^2)}{(1-M_Z^2/Q^2)^2 + M_Z^2 \Gamma_Z^2/Q^4} \biggr), 
\nn \end{align}
which depends on the (anti)quark flavor through its electric charge $Q_q$ and vector and axial couplings $v_q$ and $a_q$ to the intermediate vector boson.

\begin{figure}[t]
\includegraphics[height=40ex]{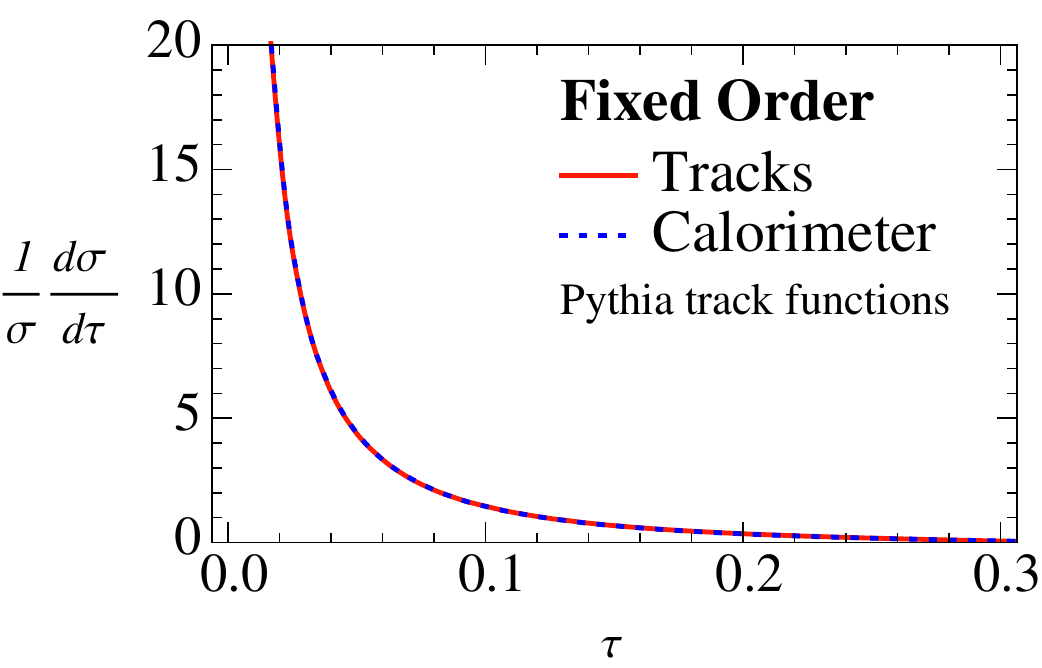} \\[-1ex]
\caption{Distributions for calorimeter and track thrust from \eq{NLO} at $\ord{\alpha_s}$. The NLO track functions are extracted from \pythia 8.150~\cite{Sjostrand:2006za,Sjostrand:2007gs} using the procedure in \Ref{Chang:2013rca}.
\vspace{-1ex}}
\label{fig:NLO}
\end{figure}

\begin{figure}[t]
\includegraphics[height=40ex]{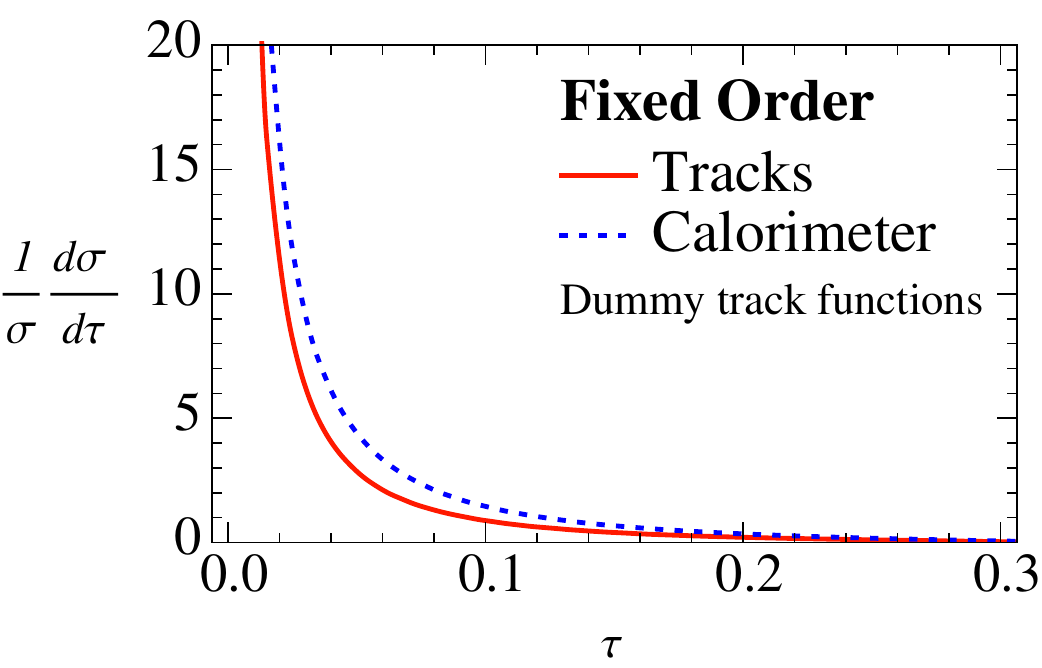} \\[-1ex]
\caption{Distributions for calorimeter and track thrust using dummy track functions.  Comparing to \fig{NLO}, we conclude that the similarity between $\tau$ and $\bar \tau$ is due to the specific form of the track function. \vspace{-1ex}}
\label{fig:NLOdummy}
\end{figure}

In \fig{NLO} we compare the calorimeter versus track thrust distributions at $\ord{\alpha_s}$, and find that they are remarkably similar. One might wonder if this small difference is a fundamental feature of \eq{NLO} or simply an accident of the specific forms of our (\pythia-based) track functions.  We can test this by calculating track thrust using the following \emph{dummy} track functions
\begin{align}
 T_q(x,\mu=M_Z) &= 30\, x^4(1-x)
 \,,  \nn \\
 T_g(x,\mu=M_Z) &= 252\, x^2 (1-x)^6
\,.\end{align}
Indeed, the difference in \fig{NLOdummy} between track and calorimeter thrust is now large. Thus, the similarity between the $\tau$ and $\bar{\tau}$ distributions has to do with the specific properties of the track function.  We will be able to achieve a better analytic understanding of why the effect of switching from calorimeter to tracks is so small in \Sec{NLL}.

\section{Factorization and Resummation of Track Thrust}
\label{sec:resum}

The thrust distribution can be divided into three regions: the peak region $(\tau  \simeq 2\Lambda_{\rm QCD}/Q)$, the tail region $(2\Lambda_{\rm QCD}/Q  \ll \tau < 1/3)$, and the far-tail region $( 1/3 \lesssim \tau \le 1/2)$.
For $\tau \simeq 0$, events are described by two narrow back-to-back jets, each carrying
about half of the center-of-mass energy.  For $\tau$ close to the kinematic endpoint $1/2$, the event is characterized by an isotropic multi-particle final state.  At $\ord{\alpha_s}$ from \Sec{NLO}, the kinematic endpoint is $1/3$ corresponding to three maximally separated jets. We therefore do not obtain a reliable description of the far-tail region.

In this paper, we are interested in properly describing the tail region of the thrust distribution, which dominantly consists of broader dijets and 3-jet events.  In this region, the dynamics is governed by three well-separated scales: the \emph{hard scale} ($\mu_H\simeq Q$) which is set by the $e^+e^-$ center-of-mass~energy $Q$, the \emph{jet scale} ($\mu_J\simeq Q\sqrt{\tau}$) which is set by the momentum of the particles transverse to thrust axis, and the \emph{soft scale} ($\mu_S\simeq Q\,\tau$) which is set by the typical energy of soft radiation between the hard jets.  When $\tau\ll 1$, there will be large hierarchies between these scales, so we will need to resum double logarithms of the form $\al_s^n \ln^m \tau$ $(m\leq 2n)$.  Because we focus on the region where $\mu_S \simeq \tau Q \gg \Lambda_{\mathrm{QCD}}$, the contribution from soft radiation is accurately described by perturbation theory, with non-perturbative effects captured by a series of power correction parameters.  We will only use the leading power correction $\bar{\Omega}^\tau_1$ in our analysis, though if were interested in describing the peak region correctly we would have to include a full non-perturbative shape function, see \sec{powercorrections}.

The leading-power factorization theorem for calorimeter thrust is well known \cite{Catani:1992ua,Korchemsky:1999kt,Fleming:2007qr,Schwartz:2007ib}:
\begin{align}\label{eq:factthcalo}
  \frac{\df \si}{\df \tau} &= \si_0 H(Q^2,\mu) \!\!\int_0^\infty\! \df k\, \df s_A\, \df s_B\, S(k,\mu) J(s_A,\mu) J(s_B,\mu)
  \nn \\ & \quad \times
\, \de\Big[\tau - \frac{1}{Q} \Big(\frac{s_A}{Q} + \frac{s_B}{Q} + k \Big) \Big]
\,.\end{align}
Here, $\sigma_0$ is the Born cross section from \eq{Bornxsec}, $H$, $J$, and $S$ are respectively the hard, jet, and soft functions, $s_{A,B}$ are the invariant mass-squareds of collinear radiation in hemispheres $A$ and $B$, and $k$ is the contribution to thrust from soft radiation.  

The goal of this section is to translate \eq{factthcalo} into a factorization theorem for track thrust.  This procedure is made straightforward by applying the matching procedure defined in \eq{matching} to the objects $S$ and $J$.  The final answer is:
\begin{align}\label{eq:factth}
  \frac{\df \si}{\df \bar \tau} &= \si_0 H(Q^2,\mu) \int_0^\infty\! \df \bar k\, \df \bar s_A\, \df \bar s_B \int_0^1\! \df x_A\, \df x_B 
  \nn \\ & \quad \times
   \bar S(\bar k,\mu) \bar J(\bar s_A,x_A,\mu) \bar J(\bar s_B,x_B,\mu)\nn\\
   & \quad \times \de\Big[\bar \tau - \frac{2}{(x_A+x_B) Q} \Big(\frac{\bar s_A}{Q} + \frac{\bar s_B}{Q} + \bar k \Big) \Big]
\,.\end{align}
We now explain each of the ingredients in this formula, with details to appear in the subsequent subsections.

The delta function in \eq{factth} comes from the form of $\bar{\tau}$ given in \eq{bartau}.  Dividing phase space into hemispheres $A$ and $B$ defined by the thrust axis, track thrust depends on the track fractions $x_{i}$, the rescaled track invariant mass-squared of collinear radiation $\bar s_i = s^{\rm tracks}_i / x_i$, and the track soft contribution $\bar k$.  The reason we are using the rescaled $\bar s_i$ (and not $s^{\rm tracks}_i$ directly) is that $\bar s_A = Q \bar{k}_A^+$ and $\bar s_B = Q \bar{k}_B^-$ directly enter the definition of track thrust.

The hard function $H(Q^2,\mu)$ is the same as for calorimeter thrust and encodes virtual effects arising from the production of the $q \bar q$ pair at the hard scale.  We give the form of $H$ in \Sec{hard}.

The track thrust soft function $\bar{S}(\bar k,\mu)$, where $\bar k = \bar k_A^+ + \bar k_B^-$, describes the contribution to track thrust due to soft parton emissions which then hadronize into tracks. At NLO, soft radiation consists of only a single gluon emission so we can simply rescale
\be
\label{eq:barkdef}
\bar k = x k,
\ee
where $x$ is the track fraction of the gluon.  This leads to a straightforward relationship between the ordinary thrust soft function and the track-based version, as discussed in \Sec{soft}. At higher orders, the expression for $\bar k$ will become more complicated. The track-based soft function also incorporates information about non-perturbative physics through power corrections, and we discuss the leading power correction ${\bar \Omega}^\tau_1$ in \Sec{powercorrections}.

The track-based jet function $\bar J(\bar s,x,\mu)$ encodes the (real and virtual) collinear radiation in each hemisphere. At NLO, a hemisphere jet consist of just two partons, so
\be
\label{eq:barsdef}
\bar s_i = \frac{x_1 x_2}{x_i} s_i,
\ee
where $x_1$ and $x_2$ are the track fractions of the two partons, $x_i$ is the track fraction of the hemisphere ($i=A,B$), and $s_i$ is the (calorimeter) invariant mass of the hemisphere. Unlike the calorimetric version, $\bar J$ depends not only on the rescaled track invariant mass $\bar{s}$ (given by \eq{barsdef} at NLO), but also on the track fraction $x$.  For this reason, the track-based jet function is considerably more complicated than the usual jet function, and requires a more complicated matching calculation, as described in \Sec{jet}. 

In order to resum logarithms, we not only need the forms of the $H$, $\bar J$, and $\bar S$, but also their anomalous dimensions.  At LL order, this means incorporating the one-loop cusp anomalous dimension to resum the Sudakov double logs.  In this paper, we incorporate NLL resummation, which includes the two-loop cusp and the one-loop non-cusp anomalous dimension terms.  Correspondingly, the running of $\alpha_s$ is consistently implemented at two loops, using the $Z$ pole value for $\alpha_s$ in \eq{alphasvalue}.  
Track thrust resummation is very similar to the calorimetric case, as discussed in \Sec{NLL} and the appendices.  The main difference is that the anomalous dimensions of $\bar J$ and $\bar S$ now depend on non-perturbative parameters.

In addition to the ingredients above, we will incorporate fixed-order non-singular corrections described in \Sec{ns}.  Following the primed counting scheme of \Ref{Abbate:2010xh}, fixed-order matching contributions are included at one order higher in the expansion in $\alpha_s$ compared to the usual (non-primed) counting.  Here we work to NLL$'$ order which incorporates all of the the ${\cal O}(\alpha_s)$ terms contained in \eq{NLO}.

\subsection{Hard Function}
\label{sec:hard}

At leading order in the electroweak interactions, the hard function is given by the square of the Wilson coefficient in the matching of the quark current from QCD onto SCET~\cite{Manohar:2003vb, Bauer:2003di},
\begin{align} \label{eq:Hfunc}
H(Q^2,\mu)
&= 1 + \frac{\alpha_s(\mu)\,C_F}{2\pi} \biggl[-\ln^2 \Bigl(\frac{Q^2}{\mu^2}\Bigr) 
\nn \\ & \quad
+ 3 \ln \Bigl(\frac{Q^2}{\mu^2}\Bigr) - 8 + \frac{7\pi^2}{6} \biggr]
\,.\end{align}
The anomalous dimension of this object is
\begin{align} \label{eq:H_RGE}
\mu \frac{\df}{\df\mu} H(Q^2, \mu) &= \gamma_H(Q^2, \mu)\, H(Q^2, \mu)
\,,\nn \\
\gamma_H(Q^2, \mu) &=
2\Gamma_\cusp[\alpha_s(\mu)] \ln\frac{Q^2}{\mu^2} + \gamma_H[\alpha_s(\mu)]
\,,\nn \\
\gamma_H[\alpha_s] &=
-\frac{3 \al_s C_F}{\pi}
\,. 
\end{align}
The cusp anomalous dimension $\Ga_\cusp$ is given in \eqs{ga_exp}{ga_coeff}. We will use the non-cusp $\gamma_H$ to perform a consistency check on our factorization theorem in \eq{factorizationconsistency}.

\subsection{Soft Function}
\label{sec:soft}

At NLO, there is only one soft gluon emission, so in order to obtain the soft function, we can simply convolve the NLO thrust soft function with the gluon track function,
\begin{align} \label{eq:scaledsoft}
  \bar S(\bar k,\mu) &= \int_0^\infty\! \df k\, S(k,\mu) \int_0^1\! \df x \, T_g(x,\mu)\, \de(\bar k - x k)
\,,
\end{align}
where we have used the relationship between the kinematics in \eq{barkdef}.  This is the simplest possible version of the matching equation in \eq{matching}.

The ordinary thrust soft function $S$ is defined through the vacuum matrix element of eikonal Wilson lines and its one-loop perturbative expression for calorimeter thrust can be obtained from \Refs{Schwartz:2007ib, Fleming:2007xt},
\begin{equation} \label{eq:calo_soft}
S(k,\mu) = \delta(k) + \frac{\alpha_s(\mu)\,C_F}{2\pi} \biggl[
-\frac{8}{\mu} \cL_1\Bigl(\frac{k}{\mu}\Bigr) + \frac{\pi^2}{6}\, \delta(k) \biggr]
\,,
\end{equation}
where the plus distributions $\cL_n$ are defined in \app{plusfunc}.

Using \eq{scaledsoft}, the corresponding track-based version $\bar S$ is given by
\begin{align} \label{eq:Sfunc}
  \bar S(\bar k,\mu) 
  & = \int_0^1\! \frac{\df x}{x}\, T_g(x,\mu) \bigg( \delta(\bar k/x) 
\nn \\ & \quad  
  + \frac{\alpha_s\,C_F}{2\pi} \biggl[
-\frac{8}{\mu} \cL_1\Bigl(\frac{\bar k}{x \mu}\Bigr) 
+ \frac{\pi^2}{6}\, \delta(\bar k/x) \biggr] \bigg)
\nn \\ &
 = \de(\bar k) + \frac{\alpha_s\,C_F}{2\pi} \biggl[
-\frac{8}{\mu} \cL_1\Bigl(\frac{\bar k}{\mu}\Bigr) + \frac{8 g^L_1}{\mu} \cL_0\Bigl(\frac{\bar k}{\mu}\Bigr) 
\nn \\ & \quad
+ \Big(\frac{\pi^2}{6} - 4 g^L_2\Big) \delta(\bar k) \biggr]
\,.\end{align}
While one naively might think that $\bar S$ would depend on the entire track function, from the rescaling properties of the plus distributions in \eq{resc_plus}, we see that only two logarithmic moments of the gluon track function appear in the soft function, namely $g^L_1$ and $g^L_2$, defined as
\begin{align}
  g^L_n(\mu) &\equiv \int_0^1\! \df x\, T_g(x,\mu) \ln^n x
\,.\end{align}

From \eq{Sfunc}, we can derive the anomalous dimension of the track soft function
\begin{align} \label{eq:S_RGE}
\mu\frac{\df}{\df\mu} \bar S(\bar k, \mu)
&= \int_0^{\bar k}\! \df \bar k'\, \gamma_{\bar S}(\bar k - \bar k', \mu)\, \bar S(\bar k', \mu)
\,,\nn \\
\gamma_{\bar S}(\bar k, \mu)
&= 4\Gamma_\cusp[\alpha_s(\mu)]\, \frac{1}{\mu} \cL_0\Big(\frac{\bar k}{\mu}\Big) 
+ \gamma_{\bar S}[\alpha_s(\mu)]\, \de(k)
\,, \nn \\
\ga_{\bar S}[\al_s] &= -\frac{4\al_s C_F}{\pi}\, g^L_1
\,. \end{align}
Interestingly, the non-cusp anomalous dimension depends on the logarithmic moment $g^L_1$ of the gluon track function.  This arises because the RG evolution sums multiple emissions, and thus the effect of the hadronization of these emissions must be exponentiated. Note that $g^L_1$ depends (weakly) on the renormalization scale $\mu$, but this effect is beyond the order that we are working.

\subsection{Leading Power Correction}
\label{sec:powercorrections}

In the tail region of the thrust distribution, non-perturbative physics is captured via power corrections.  As we will now review, the leading power correction simply acts as a shift of the soft function in \eq{Sfunc} by an amount proportional to $\lqcd$~\cite{Manohar:1994kq,Webber:1994cp,Korchemsky:1994is,Dokshitzer:1995zt}.  The amount of the shift is different for calorimeter and track thrust, but the essential formalism is the same in both cases.

Given a hadronic final state with charged hadrons $C$ and neutral hadrons $N$, we define a calorimeter measurement operator
\be
\hat{k}\ket{CN} = \sum_{i \in C,N} (|\vec{p}_i| - |\hat{t}\cdot \vec{p}_i|) \ket{CN},
\ee
where the sum runs over all hadrons in $C$ and $N$, $\hat{t}$ is the thrust axis, and $\vec{p}_i$ is the three-momentum for hadron $i$.  This operator measures the numerator of \eq{thrustdef}.  The track measurement operator is almost the same, but the sum only runs over the charged hadrons $C$.

The soft function $S$ describes the cross section to produce a measurement $k$ in the presence of back-to-back eikonal quarks.  Formally, it is defined as
\begin{align} \label{eq:Sthrustdef}
 S(k,\mu) 
&=
  \frac{1}{N_c} 
  \big\langle 0 \big|{\rm tr}\:
   \overline Y_{\!\bar n}^T Y_n\, \delta(k-\hat{k}) 
   Y_n^\dagger \overline Y_{\!\bar n}^* 
    \big| 0 \big\rangle  
 \,,
\end{align} 
where $Y_n^\dagger(0) = {\rm P} \exp\,( ig \int_0^\infty ds\, n\sdt A(ns) )$ is a (ultra)soft Wilson line in the fundamental representation,
$\overline Y_{\bar n}^\dagger$ is the analogue in the $\overline 3$ representation, and the trace is taken over color indices.

For an additive observable like thrust, the soft function factorizes into a partonic perturbative part $S^{\rm part}$ (calculated already in \eq{calo_soft}) and a non-perturbative part $S^{\rm NP}$ (also called the shape function \cite{Korchemsky:1999kt,Korchemsky:2000kp,Hoang:2007vb,Ligeti:2008ac})
\be
\label{eq:fullsoft}
S(k) = \int_0^\infty\! \df\ell \, S^{\rm part}(k - \ell) S^{\rm NP}(\ell).
\ee
In the tail region where $k\simeq Q\tau \gg \Lambda_{\rm QCD}$, we can perform an operator product expansion (OPE) on $S^{\rm NP}(\ell)$
\be
S^{\rm NP}(\ell) = \delta(\ell) - \delta'(\ell) \Omega_1^\tau + \ldots,
\ee
where the leading power correction for thrust $\Omega_1^\tau \simeq \lqcd$ is defined via the non-perturbative matrix element
\begin{align}
\label{eq:O1}
\Omega^\tau_1 = \frac{1}{N_c} \big\langle 0 \big| {\rm tr}\ \overline Y_{\bar
    n}^T(0) Y_n(0)\, \hat{k} \, Y_n^\dagger(0) \overline Y_{\bar
    n}^*(0) \big| 0 \big\rangle \, .
\end{align}
The full soft function in \eq{fullsoft} can then be approximated as a shift
\be
\label{eq:S_shift}
S(k)\simeq S^{\rm part}(k - \Omega_1^\tau) + \ORd{\frac{\alpha_s \lqcd}{k^2}} + \ORd{\frac{\lqcd^2}{k^3}}.
\ee
This in turn leads to an overall shift in the thrust distribution, whose effect is most prominent at small $\tau$.  

The formalism above applies equally well to calorimeter thrust and track thrust.  Focussing on calorimeter thrust,  the value of $\Omega_1^\tau$ must be extracted from data, since it is a fundamentally non-perturbative parameter.  Typically, one expresses $\Omega_1^\tau$ in terms of the universal power correction $\Omega_1$ \cite{Dokshitzer:1995zt,Akhoury:1995sp,Lee:2006nr}
\be
\Omega_1^\tau \equiv 2 \Omega_1,
\ee
though strictly speaking, $\Omega_1$ is only universal for measurements in the same universality class (see \Ref{Mateu:2012nk}).  Putting aside that subtlety, the analysis in \Ref{Abbate:2010xh} extracted a value of ${\Omega}_1 = 0.264 \pm 0.213\; \text{GeV}$ in the $\overline{\rm MS}$ scheme at NLL$'$ from fits to (calorimeter) thrust data.  We will therefore take a value of
\be
\Omega_1^\tau = 0.5\; \text{GeV}
\ee
for our analysis of calorimeter thrust.  As mentioned near \eq{alphasvalue}, there are strong correlations between $\alpha_s$ and $\Omega_1^\tau$, and this choice gives a reasonable (but not perfect) description of LEP data.

For track thrust, we estimate that the parameter ${\bar \Omega}^\tau_1$ entering the analogous OPE for $\bar S^{\rm NP}(\bar k)$ is given by
\be
{\bar \Omega}^\tau_1\simeq \langle x \rangle \Omega^\tau_1 = 0.3\; \text{GeV},
\ee
where we have taken the average track fraction $\langle x \rangle$ to be 0.6 \cite{Chang:2013rca}.  This approximation is only justified if the matrix element defining ${\bar \Omega}^\tau_1$ is dominated by a single gluon emission and if the gluon track function has a narrow width.  More generally, ${\bar \Omega}^\tau_1$ will encode hadronization correlations.  

We emphasize that we have applied this non-perturbative shift ${\bar \Omega}^\tau_1$ to the track-based soft function directly,
\be
\bar S(\bar k,\mu)\simeq \bar S^{\rm part}(\bar k - {\bar \Omega}_1^\tau,\mu).
\ee
Note that a shift in the track soft function $\bar S (\bar k)$ does not amount to an overall shift of the whole track thrust distribution due to the more complicated convolution structure in \eq{factth}.
Looking at \eq{scaledsoft}, we could have tried to apply the usual shift $\Omega_1^\tau$ to $S$ instead, but this would have ignored the important fact that the track function $T_g$ itself has non-perturbative power corrections.  The power correction ${\bar \Omega}^\tau_1$ includes both of these effects.  For the subleading power corrections (beyond the scope of this paper), it may or may not be preferable to separately treat the non-perturbative corrections to $S$ and $T_g$.

\subsection{Jet Function}
\label{sec:jet}

For the collinear radiation, described by the jet function, we need both the dependence on the energy fraction $x$ of the collinear tracks as well as their contribution to the rescaled hemisphere track invariant mass-squared $\bar{s}$.  The NLO jet function consists of one perturbative $q \to qg$ splitting whose branches hadronize independently.  To carry out the matching in \eq{matching}, we can use the matching coefficient $\cJ_{qq}(s,z,\mu)$ given in \Refs{Liu:2010ng,Jain:2011xz}, since the cancellation of IR divergences proceeds in an identical manner. Here, $s$ is the $q g$ invariant mass and $z$ is the momentum fraction of the final quark.  Inserting this matching coefficient into \eq{matching}, the matching calculation yields
\begin{align} \label{eq:trackjet}
  \!\!\!\bar J(\bar s,x,\mu) &= \int_0^\infty\! \df s\int_0^1\! \df z\, \frac{\cJ_{qq}(s,z,\mu)}{2(2\pi)^3}\! \nn \\
  & \quad \times  \int_0^1\! \df x_1\,\df x_2\, T_q(x_1,\mu) \, T_g(x_2,\mu)
  \nn \\ & \quad \times
  \de[x\!-\!z x_1\!-\!(1-z) x_2]\, \de\Big(\bar s - \frac{x_1 x_2}{x} s\Big)
\,,\end{align}
where we have used the kinematics in \eq{barsdef}.  The same coefficients $\cJ_{ij}(s,z,\mu)$ also appeared in the description of the fragmentation of a hadron inside a jet~\cite{Procura:2009vm,Jain:2011xz}, as they describe the perturbative splittings building up the jet radiation.

The expression for the matching coefficient is \cite{Liu:2010ng,Jain:2011xz}
\begin{widetext}
\begin{align} \label{eq:Jqqresult}
\frac{\cJ_{qq} (s,z,\mu)}{2(2\pi)^3}
& = \de(s)\, \de(1-z) 
+ \frac{\al_s(\mu) C_F}{2\pi}
\bigg\{
\frac{2}{\mu^2} \cL_1\Big(\frac{s}{\mu^2}\Big) \de(1-z) 
 + \frac{1}{\mu^2} \,\cL_0\Big(\frac{s}{\mu^2}\Big)\, (1 +z^2)  \cL_0(1-z) 
 \nn \\ & \quad
+ \de(s) \Big[(1+z^2)\,\cL_1\,(1-z) + \frac{1+z^2}{1-z} \ln z 
+ 1-z -\frac{\pi^2}{6}\de(1-z) \Big] \bigg\}
\,,
\end{align}
so evaluating \eq{trackjet}, we obtain
\begin{align}
  \bar J(\bar s,x,\mu) 
&= \bigg(\de(\bar s) + 
 \frac{\al_sC_F}{2\pi} \bigg[
\frac{2}{\mu^2} \cL_1\Big(\frac{\bar s}{\mu^2}\Big)
- \frac{2 g_1^L}{\mu^2} \cL_0\Big(\frac{\bar s}{\mu^2}\Big)  
+ \de(\bar s) \Big(g_2^L - \frac{\pi^2}{6}\Big) \bigg] \bigg) T_q(x) 
+ \frac{\al_sC_F}{2\pi} \!\int_0^1\! \df x_2 \int_0^1\!\frac{\df z}{z}
\nn \\ & \quad \times
 \bigg\{
\frac{1}{\mu^2} \cL_0\Big(\frac{\bar s}{\mu^2}\Big) (1\!+\!z^2) \cL_0(1\!-\!z) 
+ \de(\bar s) \bigg[
(1\!+\!z^2)\cL_1(1\!-\!z) + \ln\Big(\frac{xz^2}{[x\!-\!(1\!-\!z)x_2] x_2}\Big) (1\!+\!z^2) \cL_0(1\!-\!z) + 1\!-\!z \bigg]  \bigg\}  
\nn \\ & \quad \times
T_q\Big(\frac{x-(1-z)x_2}{z}\Big) T_g(x_2)
\,.\end{align}
\end{widetext}
Here we use that the track function vanishes outside the range $x \in [0,1]$ to avoid writing explicit Heaviside functions.
Unlike the soft function, the jet function depends on the full functional form of the quark and gluon track functions, and not just the logarithmic moments. To perform these integrals numerically, we used the \cuba package~\cite{Hahn:2004fe}.

The corresponding anomalous dimension is given by
\begin{align} \label{eq:J_RGE}
\mu \frac{\df}{\df \mu} \bar J(\bar s, x, \mu) &= \int_0^{\bar s}\! \df \bar s'\, \gamma_{\bar J}(\bar s-\bar s',\mu)\, \bar J(\bar s',x, \mu)
\,,\nn\\
\gamma_{\bar J}(\bar s, \mu)
&= -2 \Gamma_\cusp[\alpha_s(\mu)] \frac{1}{\mu^2}\cL_0\Bigl(\frac{\bar s}{\mu^2}\Bigr) \nn\\
& \quad+ \gamma_{\bar J}[\alpha_s(\mu)]\delta(\bar s)
, \nn \\
\ga_{\bar J}[\al_s] &= \frac{\al_s C_F}{\pi} \Big(2g_1^L + \frac{3}{2} \Big)
\,.\end{align}
Note that the evolution only affects $\bar s$ and not $x$. As for the soft function, the logarithmic moment of the gluon track function $g_1^L$ contributes to the non-cusp anomalous dimension.

\subsection{Resummation}
\label{sec:resumdetails}

In the effective field theory approach we follow here, the resummation of large double logarithms $\al_s^n \ln^m \tau$ $(m\leq 2n)$ is achieved by evaluating the hard, jet, and soft functions at their natural scales $\mu_H$, $\mu_J$, and $\mu_S$ where they contain no large logarithms, and running them to a common scale $\mu$ using their respective RG equations.  

These RG evolution kernels were implicit in the cross section in \eq{factth} and are given in \app{running}. Explicitly including them,
\begin{align} \label{eq:evol_xsec}
\frac{\df\sigma}{\df \bar\tau}
&= 
\si_0\, H(Q^2, \mu_H)\, U_H(Q^2, \mu_H, \mu) 
\nn\\ 
& \quad\times \int\!\df {\bar s}_A\, \df {\bar s}_A'\, {\bar J}({\bar s}_A - {\bar s}_A', \mu_J)\, U_{\bar J}({\bar s}_A', \mu_J, \mu)
\nn \\
& \quad \times \int\!\df {\bar s}_B\, \df {\bar s}_B'\, {\bar J}({\bar s}_B - {\bar s}_B', \mu_J)\, U_{\bar J}({\bar s}_B', \mu_J, \mu)
\nn \\ 
& \quad \times
\int\!\df {\bar k}\, \df {\bar k}'\, {\bar S}({\bar k}-{\bar k}',\mu_S)\, U_{\bar S}({\bar k}', \mu_S, \mu)
\nn \\
& \quad \times \de\Big[\bar \tau \!-\! \frac{2}{(x_A\!+\!x_B) Q} \Big(\frac{\bar s_A}{Q} \!+\! \frac{\bar s_B}{Q} \!+\! \bar k \Big) \Big]
\,.
\end{align}

Consistency of the factorization theorem requires that the cross section is $\mu$-independent at the order that we are working, implying a cancellation between the anomalous dimensions.
For the cusp anomalous dimension, this cancellation is the same as for calorimeter thrust.  For the non-cusp pieces from \eqss{H_RGE}{S_RGE}{J_RGE}, there are additional terms involving $g^L_1$ in $\ga_{\bar S}$ and $\ga_{\bar J}$, but they cancel in the sum
\be
\label{eq:factorizationconsistency}
\gamma_H^{\rm non-cusp} + 2 \gamma_{\bar{J}}^{\rm non-cusp} + \gamma_{\bar{S}}^{\rm non-cusp} = 0
\, , 
\ee
to fullfil consistency requirements.
 
An important question is the choice of scales $\mu_H$, $\mu_J$, and $\mu_S$ to use in this formula.  While our focus is on the tail region of the thrust distribution, where $\mu_H \simeq Q$, $\mu_J \simeq \sqrt{\tau} Q$ and $\mu_S \simeq \tau Q$, we do want our formulas to be accurate for all values of $\tau$.  Since there are three distinct kinematic regions characterizing the thrust distribution, the resummation of the logarithms of $\tau$ must be handled in different ways.  A smooth transition between the three regions is achieved through profile functions~\cite{Ligeti:2008ac,Abbate:2010xh} as described in \app{profiles}. Our choice of the profile parameters is such that resummation is turned off at $\bar \tau \simeq 1/3$, which is the $\mathcal{O}(\alpha_s)$ endpoint from \Sec{NLO}.  (This is in contrast to the higher-order calculation in \Ref{Abbate:2010xh,Becher:2008cf} where the resummation is only turned off at the true endpoint $\tau \simeq 1/2$.) 

For the plots in \Sec{results}, we calculate the cumulative version of \eq{evol_xsec}
\be
\label{eq:cumulativeTrack}
\bar \Sigma(\bar \tau^c) \equiv  \int_0^{\bar \tau^c}\!\!\! \df \bar \tau\, \frac{\df \si}{\df \bar \tau}
\ee
at NLL$'$ using the scales $\mu_H$, $\mu_J$, and $\mu_S$ set by the value of $\bar{\tau}^c$.  We then take the numerical derivative of $\bar \Sigma(\bar \tau^c)$ to find the track thrust distribution (see \Ref{Abbate:2010xh} for a discussion of alternative choices).  This derivative picks up both the explicit $\bar{\tau}$-dependence as well as the implicit $\bar{\tau}$-dependence of our scale choice for $\mu_H$, $\mu_J$, and $\mu_S$.  The differential version in \eq{evol_xsec} misses the latter contribution, though it is a small effect.

\subsection{Non-Singular Contribution}
\label{sec:ns}

The factorization theorem in \eqs{factth}{evol_xsec} includes all the terms in the track thrust distribution that are singular in $\tau$ as $\tau \to 0$.  There is an additional non-singular contribution of $\ord{\tau}$, which is thus important in the endpoint region. This contribution needs to be included to have our distribution formally accurate to $\ord{\alpha_s}$ and is the last step in attaining NLL$'$ accuracy.

We can extract the non-singular corrections by subtracting the singular terms (obtained from setting $\mu_H = \mu_J = \mu_S = \mu$ in \eq{factth}) from the fixed-order $\ord{\alpha_s}$ cross section in \eq{NLO}.  At the level of the cumulative cross section in \eq{cumulativeTrack}
\begin{equation}
  \bar \Sigma_{\rm ns}(\bar \tau^c) = \bar \Sigma_{\rm FO}(\bar \tau^c) - \bar \Sigma_{\rm sing}(\bar \tau^c)
\,.\end{equation}

\begin{figure} [t]
\includegraphics[height=37ex]{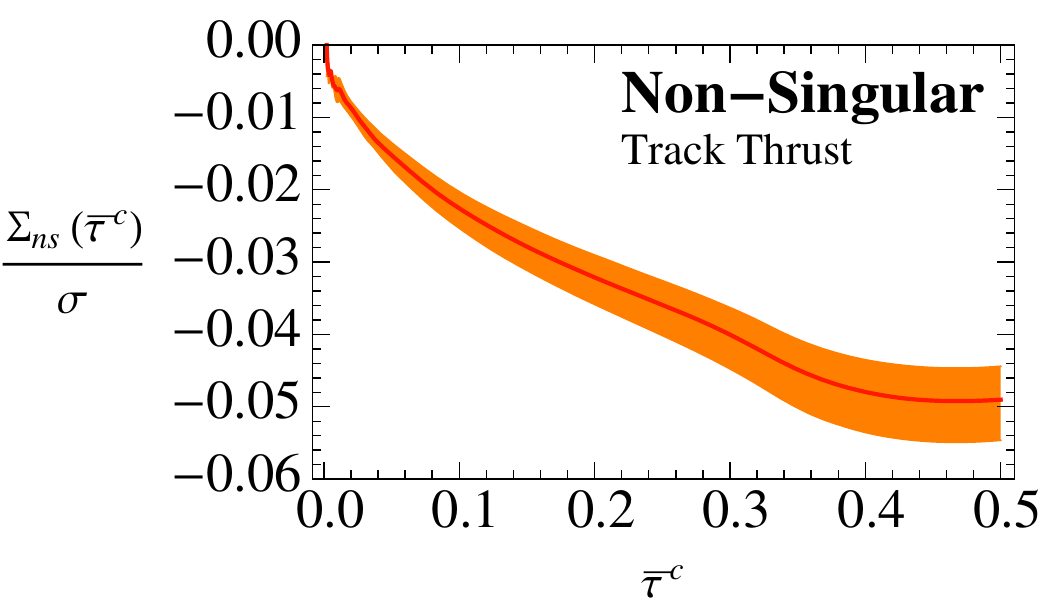} \\[-1ex]
\caption{Non-singular contribution to the normalized cumulative thrust distribution at $\ord{\alpha_s}$.  The central value corresponds to $\mu = M_Z$, with the uncertainty bands from varying $\mu \in  [M_Z/2, 2 M_Z]$.}
\label{fig:ns}
\end{figure}

Our extraction of $\bar \Sigma_{\rm ns}(\bar \tau^c)$ is shown in \fig{ns}. The fact that $\bar \Sigma_{\rm ns}(\bar \tau^c=0)=0$ provides another consistency check of our formalism, showing that our factorization formula successfully reproduces the singular part of the $\ord{\al_s}$ cross section.  We use $\mu = M_Z$ as the central value for extracting $\bar \Sigma_{\rm ns}(\bar \tau^c)$, and estimate perturbative uncertainties by varying $\mu$ between $M_Z/2$ and $2 M_Z$.

\section{Simplifications at NLL}
\label{sec:NLL}

In both the LEP data in \fig{rawdata} and the fixed-order calculation in \fig{NLO}, we saw a remarkable similarity between the calorimeter and track thrust distributions.  We will now try to understand this from our resummed calculation by looking at the leading effect of switching to tracks.

The first non-trivial order in the resummed distribution is NLL.  This consists of evaluating \eq{evol_xsec} with only the leading order hard, jet, and soft functions, but including the subleading evolution kernels. Using the solutions to the RG equations in \app{running}, the NLL cumulative distribution is
\begin{align} \label{eq:NLLfullform}
\bar \Sigma(\bar \tau^c) & = \si_0\, e^{K_H} \Bigl(\frac{Q^2}{\mu_H^2}\Bigr)^{\eta_H} 
 \frac{e^{K_{\bar S} -\gamma_E\, \eta_{\bar S}}}{\Gamma(1+\eta_{\bar S})}\, \Big(\frac{Q \bar \tau^c}{\mu_S}\Big)^{\eta_{\bar S}}
 \\ & \times 
 \int\! \df x_A\, \df x_B\, T_q(x_A,\mu_J)\, T_q(x_B,\mu_J)\, \Big(\frac{x_A\!+\!x_B}{2}\Big)^{\eta_{\bar S}}
 \!,\nn
\end{align}
where $\gamma_E$ is Euler's constant, and we have chosen to evolve the hard and soft scales to the jet scale $\mu_J$.  The functions $K_H(\mu_H, \mu_J)$, $\eta_H(\mu_H, \mu_J)$, $K_{\bar S}(\mu_S, \mu_J)$ and $\eta_{\bar S}(\mu_S, \mu_J)$ are given in \app{running} and depend on our choice for $\mu_H$, $\mu_J$, and $\mu_S$, which we discuss below. Note that this expression contains an explicit dependence on the quark track functions $T_q$ since they appear in the tree-level jet functions. \eq{NLLfullform} contains only the information needed at NLL accuracy, and therefore does not include the leading hadronization power correction or non-singular contributions. 

There are various steps we can take to simplify the expression in \eq{NLLfullform}. We first consider the scales $\mu_H$, $\mu_J$, and $\mu_S$.  In \sec{resumdetails}, we advocated the use of the profile functions in \app{profiles} to achieve a smooth transition between the different regions of the thrust distribution. Here, we simplify our choice of natural scales to obtain a more illuminating analytic formula:
\be
\label{eq:naturalscales}
\mu_H = Q, \qquad \mu_J = \sqrt{3\bar \tau^c} Q,  \qquad \mu_S = 3\bar \tau^c Q
\,.
\ee
This choice has still the effect of turning off the resummation at $\bar \tau^c = 1/3$.

Second, we can simplify the dependence on the two quark track functions.  Defining
\begin{equation} \label{eq:q1L}
\qL (\mu) \equiv \int\! \df x_A\, \df x_B\, T_q(x_A,\mu)\, T_q(x_B,\mu)\, \ln \!\Big(\frac{x_A+x_B}{2} \Big)
\,,
\end{equation}
it is helpful to use the approximation
\begin{equation} 
\label{eq:expql}
\int\! \df x_A\, \df x_B\, T_q(x_A)\, T_q(x_B)\, \Big(\frac{x_A+x_B}{2}\Big)^{\eta_{\bar S}}
\approx \exp(\qL \eta_{\bar S}) 
\, .
\end{equation}
This is formally justified only for $\eta_{\bar S} \ll 1$, but for the (\pythia-based) track functions, the error is only a few percent even for $\eta_{\bar S} = 1$.  By contrast, using a linear (as opposed to exponential) approximation in \eq{expql} would yield a $\simeq 20\%$ error at $\eta_{\bar S} =1$.

Finally, because the only difference between the NLL evolution kernels for calorimeter thrust and track thrust appears in the non-cusp anomalous dimensions, we can write the track thrust cumulative $\bar \Sigma$ in terms of the calorimeter thrust cumulative $\Sigma$ as
\begin{equation} 
\bar \Sigma(\bar \tau^c) = \Sigma(\bar \tau^c) \exp({K_{\bar S} - K_{S}}) \exp(\qL \eta_{\bar S})
\,.\end{equation}
From \eq{Keta}, we find that the difference between $K_{\bar S}$ and $K_S$ is
\begin{align}
K_{\bar S}(\mu_S, \mu_J) - K_{S}(\mu_S, \mu_J) & = \frac{8C_F g_1^L}{\beta_0} \ln \frac{\al_s(\mu_J)}{\al_s(\mu_S)} 
\nn \\ &
\approx  \frac{4 \al_s C_F}{\pi}\, g_1^L \ln \frac{\mu_S}{\mu_J}
\nn \\ &
= \frac{2 \al_s C_F}{\pi}\, g_1^L \ln (3\bar\tau^c)
\,.\end{align}
Here we used the running of $\al_s$ to obtain the second line, and inserted the natural scales from \eq{naturalscales} in the last step. (Since we only kept the leading term in $\al_s$, different choices for the scale of $\al_s$ correspond to effects beyond the order we are working.)
Similarly, we find that $\eta_{\bar S}$ is given by
\begin{align}
 \eta_{\bar S}(\mu_S, \mu_J) & = -\frac{8C_F}{\beta_0} \ln \frac{\al_s(\mu_J)}{\al_s(\mu_S)} 
\nn \\ &
\approx - \frac{2 \al_s C_F}{\pi}\,  \ln (3\bar\tau^c)
\,.\end{align}
This leads to
\begin{equation}
\label{eq:moreapproxNLLcumulative}
\bar \Sigma(\bar \tau^c)
 \approx  \Sigma(\bar \tau^c)  \exp\bigg[\frac{2 \al_s C_F}{\pi} (g_1^L - \qL) \ln (3\bar \tau^c) \bigg]
\,,\end{equation}
as anticipated in \eq{approxNLLcumulative}.

Based on \eq{moreapproxNLLcumulative}, we now have a better understanding of why track thrust and calorimeter thrust are so similar.  At NLL order, the difference between the cumulative distributions for track and calorimeter thrust is basically given by an exponential factor. However, this factor depends on $g_1^L$ and $\qL$, which happen to be nearly equal for the track functions extracted from \pythia. 
For concreteness, we evaluate $g_1^L$ and $\qL$ at the scale $\mu \simeq 20$ GeV, though any choice of scale between $\mu_S$ and $\mu_J$ is acceptable at this order. We find
\be \label{eq:logmoments}
g_1^L \simeq -0.52, \quad \qL \in [-0.49, -0.54]
\,,\ee
where the range corresponds to the variation between different quark flavors.  This leads to a cancellation in \eq{moreapproxNLLcumulative}, which is responsible for the similarity between the calorimeter and track thrust distributions. These parameters have only a mild $\mu$-dependence, and the partial cancellation between $g_1^L$ and $\qL$ persists over a wide range of scales.

\section{Numerical Results}
\label{sec:results}

With all of the ingredients for the track thrust distribution in place, we now show numerical results as we increase the accuracy of our calculation.  In all cases, we show normalized cross sections $(1/\sigma) (d \sigma/ d \tau)$, and use our (\pythia-based) NLO track functions as input.

\begin{figure}
\includegraphics[height=40ex]{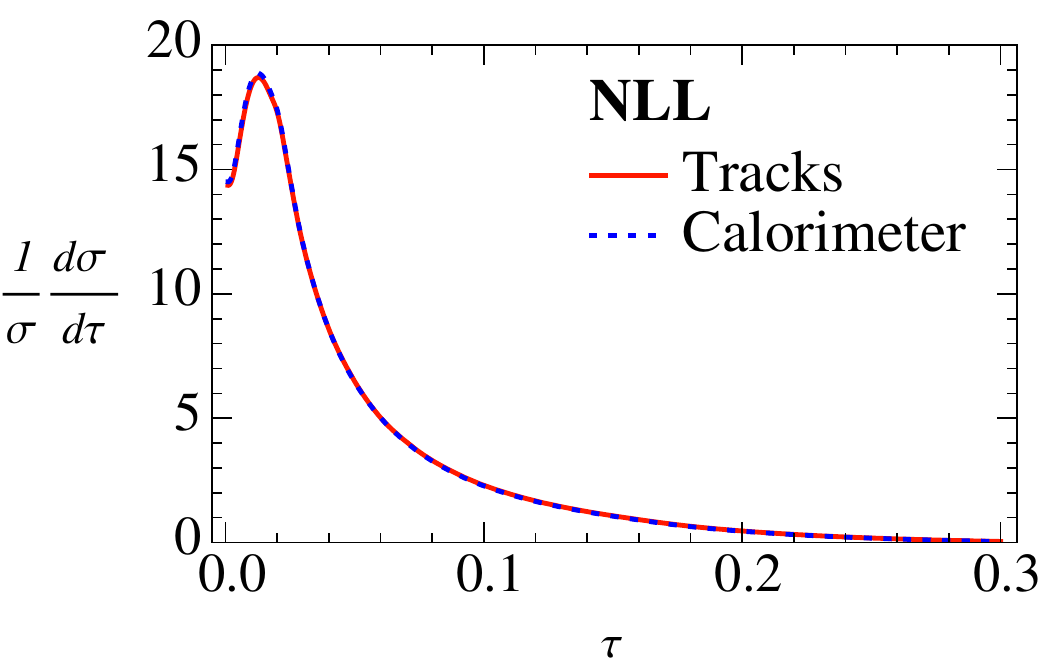} \\[-1ex]
\caption{Track thrust and calorimeter thrust at NLL.  As explained in \Sec{NLL}, these distributions are remarkably similar.}
\label{fig:nll}
\end{figure}

In \fig{nll}, we show the NLL result from \eq{NLLfullform} for calorimeter and track thrust.  Here we use the central values for the canonical running scales described in \app{profiles}.  As argued in \sec{NLL}, the difference between calorimeter and track thrust is very small at NLL order, and is in fact barely visible on this plot.  

\begin{figure}
\includegraphics[height=40ex]{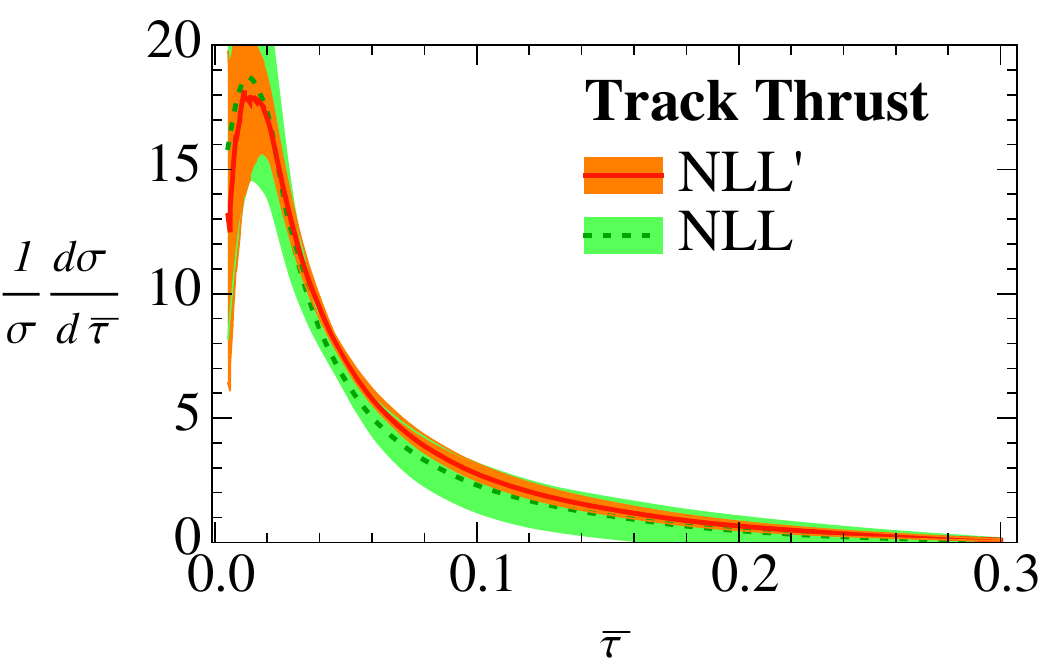} \\[-1ex]
\caption{Track thrust distribution going from NLL to NLL$'$.  The bands encode perturbative uncertainties from RG scale variations, but not uncertainties in $\alpha_s$ or the track functions themselves. \vspace{-1ex}}
\label{fig:conv}
\end{figure}

To achieve NLL$'$ accuracy, we have to take into account higher-order terms in $H$, $\bar J$, and $\bar S$ in \eq{evol_xsec}, as well as the non-singular terms from \Sec{ns}.  The result of going from NLL to NLL$'$ is shown in \fig{conv}, which compares the track thrust distributions in the peak and tail regions.  The inclusion of the one-loop corrections to the hard, jet, and soft functions at NLL$'$ reduces the purely perturbative uncertainty bands coming from scale variations.  Note that this uncertainty estimate does not include the uncertainty associated with the value of $\al_s(M_Z)$ or with the input track functions.

\begin{figure}
\includegraphics[height=40ex]{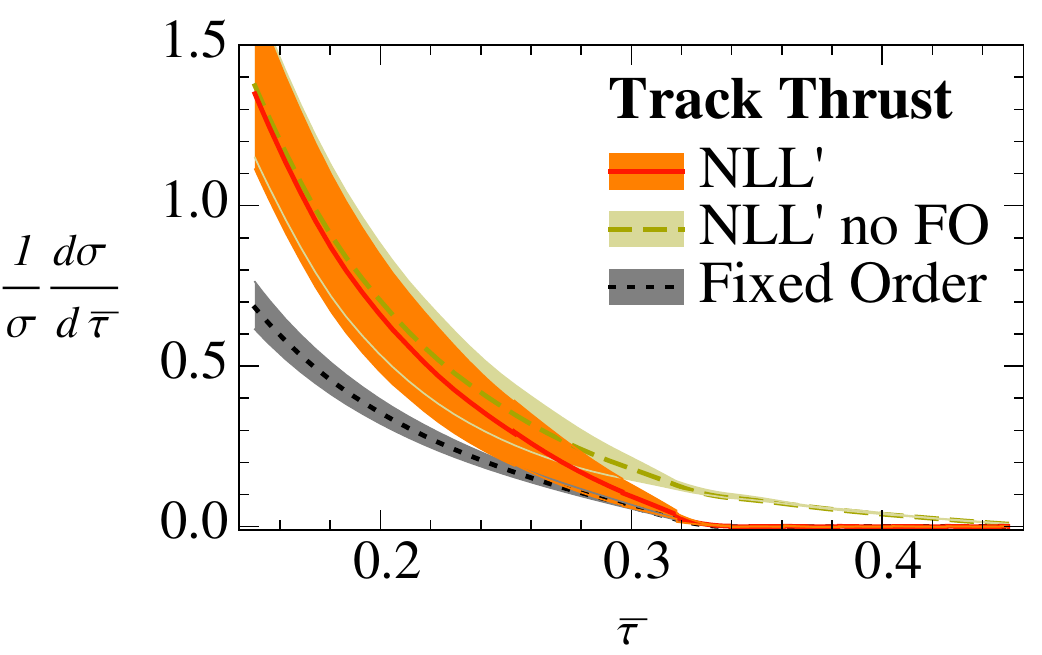} \\[-1ex]
\caption{Track thrust distribution in the tail and far-tail regions, illustrating the effect of including the non-singular contribution at NLL$'$ order.  The full NLL$'$ distribution interpolates between the resummed and fixed-order results. \vspace{-1ex}}
\label{fig:tail}
\end{figure}

The effect of the non-singular terms on the tail and far-tail regions are highlighted in \fig{tail}.  The inclusion of these terms guarantees that the cross section merges with the $\ord{\alpha_s}$ fixed-order result in the region where the resummation is no longer important.  It also ensures that the cross section vanishes beyond the $\mathcal{O}(\alpha_s)$ kinematic endpoint $\tau=1/3$.  (For this to happen, it is crucial that the profile functions in \app{profiles} turn off the resummation at the endpoint.) As desired, the full NLL$'$ distribution interpolates between the NLL$'$ result (without non-singular terms) at small $\tau$ and the fixed-order result at large $\tau$.

\begin{figure}
\includegraphics[height=40ex]{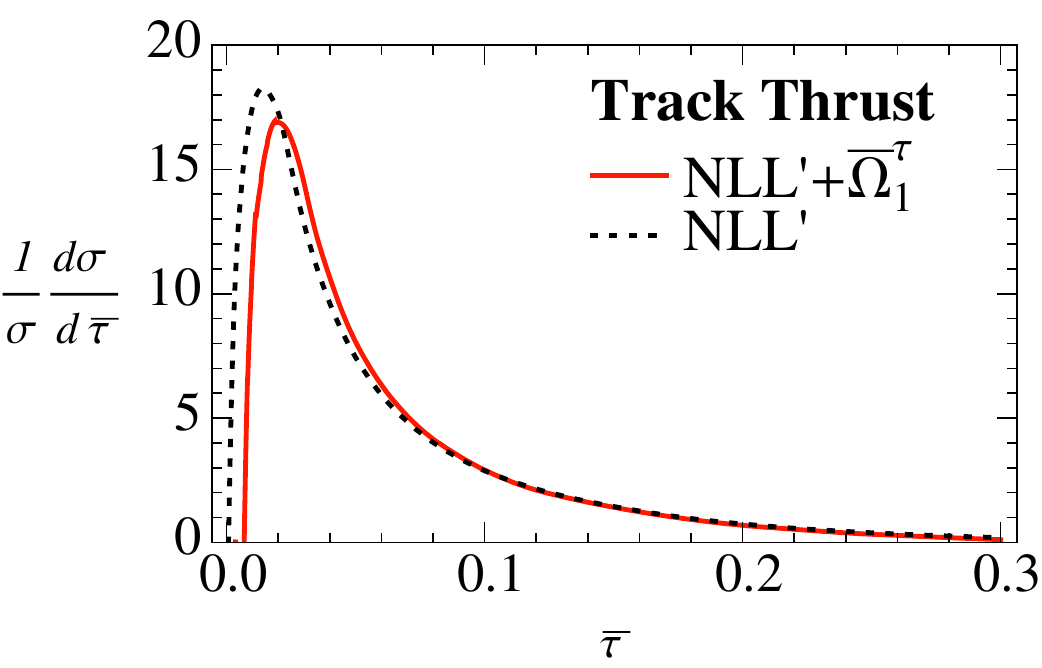} \\[-1ex]
\caption{Track thrust at NLL$'$ adding the leading power correction.}
\label{fig:nllprimepluspower}
\end{figure}

In \fig{nllprimepluspower}, we augment the NLL$'$ results with the leading power correction $\bar \Omega^\tau_1$.  For track thrust, the dominant effect of $\bar \Omega^\tau_1$ is a shift, though there are important effects in the peak region which do not amount to a shift.   (For the calorimeter thrust distribution, the only effect of $\Omega^\tau_1$ is to shift the distribution.)  Note, however, that the peak region is also sensitive to higher-order power corrections which we have not included.  The comparison between calorimeter and track thrust with the leading power correction is shown in \fig{summary}.

\begin{figure}
\includegraphics[height=40ex]{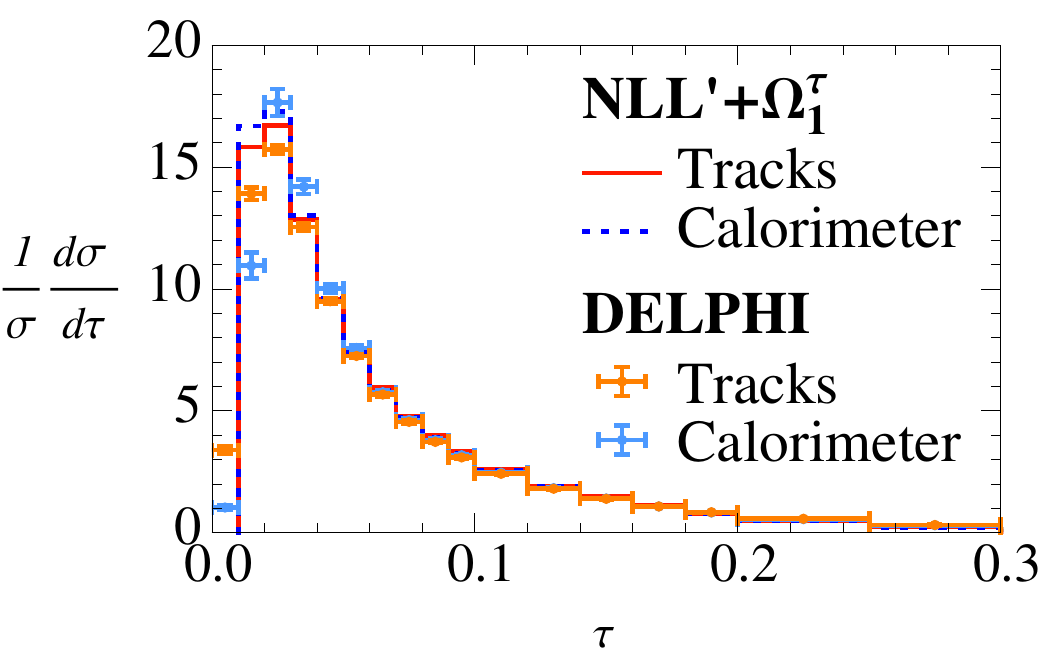} \\[-1ex]
\caption{Comparison of analytical predictions with DELPHI data for both track and calorimeter thrust distributions. There is good qualitative and quantitative agreement in the tail region, though as shown in \fig{summary}, the theoretical uncertainties at NLL$'$ are larger than the experimental ones. \vspace{-1ex}}
\label{fig:data}
\end{figure}

In \fig{data} we superimpose our theoretical predictions for the calorimeter and track thrust distributions with experimental data from the DELPHI collaboration.  At NLL$'$ order with the leading power correction $\bar \Omega^\tau_1$, the agreement is quite good, though we emphasize that we chose values of $\alpha_s$ and $\bar \Omega^\tau_1$ to ensure reasonable agreement with the calorimeter thrust data.  We show the effect of scale uncertainties in \fig{summary}, which are in general larger than the experimental uncertainties, motivating future studies of track thrust with higher orders of resummation and more accurate fixed-order corrections.

\begin{figure}
\includegraphics[height=40ex]{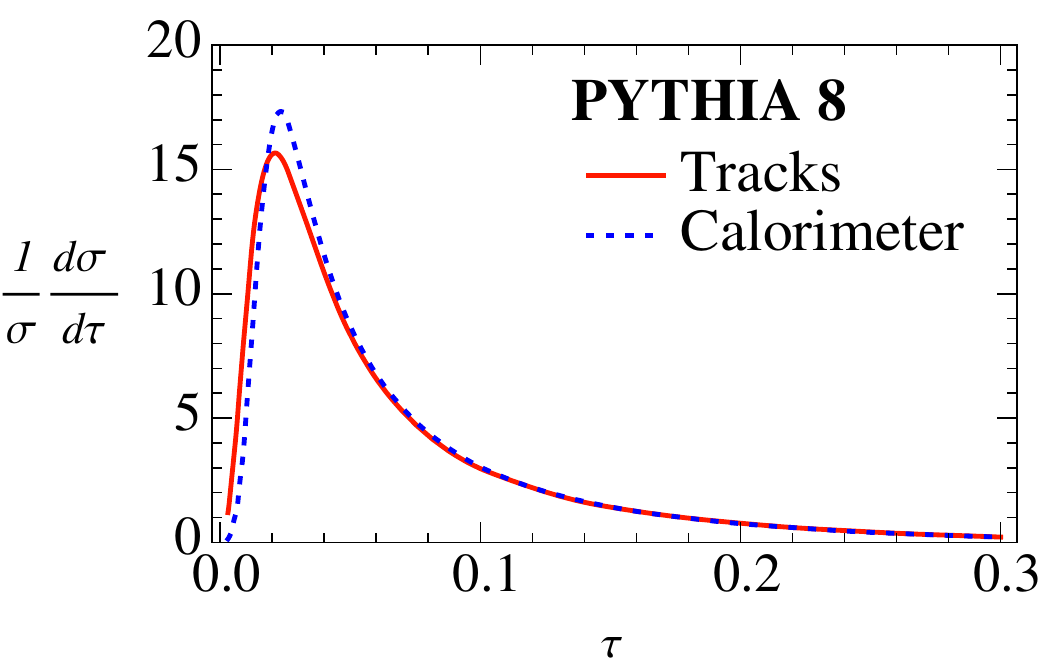} \\[-1ex]
\caption{Calorimeter and track thrust distributions obtained from \pythia~8.  Apart from deviations in the peak region due to higher-order non-perturbative corrections, these agree well with our NLL$'$ calculation after the leading power correction is included (compare to \fig{summary}).
\vspace{-1ex}}
\label{fig:pythia}
\end{figure}

As a final cross check of our analysis, we show the calorimeter and track thrust distributions from \pythia in \fig{pythia}.  Since \pythia\ has been tuned to LEP data, it agrees well with the DELPHI measurements.  There is good agreement between \pythia\ and our NLL$'$ result in the tail region, but there are difference in the peak region due to the fact that \pythia\ includes an estimate of the full non-perturbative corrections, whereas we only include the leading power correction.  Future track thrust calculations could use a full non-perturbative shape function for better modeling of the $\bar{\tau} \simeq 0$ region.

\section{Discussion}
\label{sec:discussion}

In this paper, we have presented the first calculation of track thrust in perturbative QCD.  Our result is accurate to $\ord{\alpha_s}$ in a fixed-order expansion while also including NLL resummation, i.e.~NLL$'$ order.  By incorporating both track functions and the leading power correction, we have accounted for the dominant non-perturbative effects that determine the track thrust distribution. Our result is in good agreement with track thrust measurements performed at ALEPH and DELPHI.

One feature seen in the data is a remarkable similarity between the calorimeter thrust and track thrust distributions.  At NLL, we traced this feature to a partial cancellation between two non-perturbative parameters---one associated with the gluon track function $g_1^L$, and one associated with pairs of quark track functions $\qL$.  We conjecture that a similar cancellation should be present in most (if not all) dimensionless track-based observables.  This should be relatively straightforward to prove for $e^+ e^-$ dijet event shapes with a thrust-like factorization theorem, but is likely to persist for more general track-based observables, including jet shapes relevant for the LHC such as $N$-subjettiness ratios \cite{Thaler:2010tr,Thaler:2011gf} or energy correlation functions ratios \cite{Larkoski:2013eya}.  It is worth further study to understand whether this partial cancellation is just an accident or reflects some deeper property of track functions. Crucially, we have seen that neither higher-order terms at NLL$'$ nor the leading power correction qualitatively spoil the similarity.

The track functions were originally designed to describe the energy fraction of a parton carried by tracks (i.e.~the large component of the light-cone momentum).  Track thrust essentially measures the  small component of the light-cone momentum carried by tracks, so it is perhaps surprising that the same track functions can be used in this context.  The reason this works is that the track thrust distribution can be thought of as arising from multiple gluon emissions, each of which carries its own track function.  Just as multiple emissions can be exponentiated in the case of calorimeter thrust, multiple emissions with track functions can also be exponentiated.  In our calculation, this shows up in the fact that the anomalous dimension of the soft and jet functions depend on the logarithmic moment of the gluon track function $g_1^L$.  We are confident that similar techniques could be applied to any track-based observable, as long as the calorimetric version of that observable has a valid factorization theorem.  This motivates future experimental and theoretical studies of track-based observables.

\begin{acknowledgments}
We thank Iain Stewart for discussions. H.C.~and W.W.~are supported by the U.S. Department of Energy (DOE) grant DE-FG02-90ER-40546.  M.P.~acknowledges support  by the ``Innovations- und Kooperationsprojekt C-13'' of the Schweizerische Universit\"atskonferenz SUK/CRUS and by the Swiss National Science Foundation. J.T. is supported by the DOE under cooperative research agreement DE-FG02-05ER-41360 and under the DOE Early Career research program DE-FG02-11ER-41741.
\end{acknowledgments}

\appendix 

\section{Resummation}
\label{app:running}

For the NLL$'$ distribution in \eq{evol_xsec}, we need expressions for the evolution kernels.  Apart from the non-perturbative parameter $g^L_1$, the evolution kernels are the same between calorimeter thrust and track thrust, and governed by the relevant RGEs given in \Sec{resum}.

The RGE for the hard function in \eq{H_RGE} leads to the evolution
\begin{align}
H(Q^2, \mu) &= H(Q^2, \mu_0)\, U_H(Q^2, \mu_0, \mu)
\,,\nn \\
U_H(Q^2, \mu_0, \mu)
&= e^{K_H(\mu_0, \mu)} \Bigl(\frac{Q^2}{\mu_0^2}\Bigr)^{\eta_H(\mu_0, \mu)}
\,,\nn \\
K_H(\mu_0,\mu) &= -4K_\Gamma(\mu_0,\mu) + K_{\gamma_H}(\mu_0,\mu)
\,, \nn \\
\eta_H(\mu_0,\mu) &= 2\eta_\Gamma(\mu_0,\mu)
\,,\end{align}
where $K_\Gamma(\mu_0, \mu)$, $\eta_\Gamma(\mu_0, \mu)$ and $K_\gamma$ are given below in \eq{Keta_def}.
Similarly, the RGE for the jet function in \eq{J_RGE} leads to the evolution
\begin{align} 
\bar J(\bar s,x,\mu) & =  \int_0^{\bar s}\! \df \bar s'\, U_{\bar J}(\bar s - \bar s',\mu_0, \mu)\, \bar J(\bar s',x,\mu_0)
\,, \nn \\
U_{\bar J}(\bar s, \mu_0, \mu) &= \frac{e^{K_{\bar J} -\gamma_E\, \eta_{\bar J}}}{\Gamma(1+\eta_{\bar J})}\,
\biggl[\frac{\eta_{\bar J}}{\mu_0^2} \cL^{\eta_{\bar J}} \Bigl( \frac{\bar s}{\mu_0^2} \Bigr) + \delta(\bar s) \biggr]
\,, \nn \\
K_{\bar J}(\mu_0,\mu) &= 4 K_\Gamma(\mu_0,\mu) + K_{\gamma_{\bar J}}(\mu_0,\mu)
\,, \nn \\
\eta_{\bar J}(\mu_0,\mu) &= -2\eta_{\Gamma}(\mu_0,\mu)
\,.\end{align}
The function $K_{\gamma_{\bar J}}$ contains the contribution from the non-perturbative parameter $g^L_1$ to the non-cusp anomalous dimension $\gamma_{\bar J}[\al_s]$.
Finally, the RGE for the soft function in \eq{S_RGE} leads to the evolution
\begin{align} 
\bar S(\bar k,\mu) & =  \int_0^{\bar k}\! \df {\bar k}'\, U_{\bar S}(\bar k - \bar k',\mu_0,\mu)\, {\bar S}(\bar k',\mu_0)
\,, \nn \\
U_{\bar S}(\bar k, \mu_0, \mu) & = \frac{e^{K_{\bar S} -\gamma_E\, \eta_{\bar S}}}{\Gamma(1+\eta_{\bar S})}\,
\biggl[\frac{\eta_{\bar S}}{\mu_0} \cL^{\eta_{\bar S}} \Big( \frac{\bar k}{\mu_0} \Big) + \delta(\bar k) \biggr]
\,, \nn \\
K_{\bar S}(\mu_0,\mu) &= - 4 K_\Gamma(\mu_0,\mu) + K_{\gamma_{\bar S}}(\mu_0,\mu)
\,, \nn \\
\eta_{\bar S}(\mu_0,\mu) &= 4\eta_\Gamma(\mu_0,\mu)
\,.\end{align}
Here, $K_{\gamma_{\bar S}}$ contains the contribution from $g^L_1$ to $\gamma_{\bar S}[\al_s]$.

The functions $K_\Gamma(\mu_0, \mu)$, $\eta_\Gamma(\mu_0, \mu)$, and $K_{\gamma_x}(\mu_0, \mu)$ in the above RGE solutions are defined as
\begin{align} \label{eq:Keta_def}
K_\Gamma^i(\mu_0, \mu)
& = \int_{\alpha_s(\mu_0)}^{\alpha_s(\mu)}\!\frac{\df\alpha_s}{\beta(\alpha_s)}\,
\Gamma_\cusp^i(\alpha_s) \int_{\alpha_s(\mu_0)}^{\alpha_s} \frac{\df \alpha_s'}{\beta(\alpha_s')}
\,,\nn \\
\eta_\Gamma^i(\mu_0, \mu)
&= \int_{\alpha_s(\mu_0)}^{\alpha_s(\mu)}\!\frac{\df\alpha_s}{\beta(\alpha_s)}\, \Gamma_\cusp^i(\alpha_s)
\,,\nn \\
K_{\gamma_x}(\mu_0, \mu)
& = \int_{\alpha_s(\mu_0)}^{\alpha_s(\mu)}\!\frac{\df\alpha_s}{\beta(\alpha_s)}\, \gamma_x(\alpha_s)
\,,\end{align}
and their explicit expressions at NLL order are
\begin{align} \label{eq:Keta}
K_\Gamma(\mu_0, \mu) &= -\frac{\Gamma_0}{4\beta_0^2}\,
\biggl\{ \frac{4\pi}{\alpha_s(\mu_0)}\, \Bigl(1 - \frac{1}{r} - \ln r\Bigr)
\nn \\ & \quad
   + \biggl(\frac{\Gamma_1 }{\Gamma_0 } - \frac{\beta_1}{\beta_0}\biggr) (1-r+\ln r)
   + \frac{\beta_1}{2\beta_0} \ln^2 r \bigg\}
\,, \nn\\
\eta_\Gamma(\mu_0, \mu) &=
 - \frac{\Gamma_0}{2\beta_0}\, \biggl[ \ln r
 + \frac{\alpha_s(\mu_0)}{4\pi}\, \biggl(\frac{\Gamma_1 }{\Gamma_0 }
 - \frac{\beta_1}{\beta_0}\biggr)(r-1)\biggr]
\,, \nn\\
K_{\gamma_x}(\mu_0, \mu) &=
 - \frac{\gamma_{x,0}}{2\beta_0}\, \ln r
\,.\end{align}
Here $r = \alpha_s(\mu)/\alpha_s(\mu_0)$, and $\beta_i$, $\Ga_i,$  and $\ga_{x,i}$ are the coefficients of the $\beta$-function, the cusp, and the non-cusp anomalous dimensions in their $\alpha_s$ expansion,
\begin{align} \label{eq:ga_exp}
\beta(\alpha_s) &=
- 2 \alpha_s \sum_{n=0}^\infty \beta_n\Bigl(\frac{\alpha_s}{4\pi}\Bigr)^{n+1}
\,, \nn \\
\Gamma_\cusp(\alpha_s) &= 
\sum_{n=0}^\infty \Gamma_n \Bigl(\frac{\alpha_s}{4\pi}\Bigr)^{n+1}
\,, \nn \\
\gamma_x(\alpha_s) &= 
\sum_{n=0}^\infty \gamma_{x,n} \Bigl(\frac{\alpha_s}{4\pi}\Bigr)^{n+1}
\,.\end{align}
At NLL$'$ order, we only need the first two coefficients of $\beta(\alpha_s)$ and $\Gamma_\cusp(\alpha_s)$, which are 
\begin{align} \label{eq:ga_coeff}
\beta_0 &= \frac{11}{3}\,C_A -\frac{4}{3}\,T_F\,n_f
\,,\nn\\
\beta_1 &= \frac{34}{3}\,C_A^2  - \Bigl(\frac{20}{3}\,C_A\, + 4 C_F\Bigr)\, T_F\,n_f
\,,\nn\\
\Gamma_0 &= 4C_F
\,,\nn\\
\Gamma_1 &= 4C_F \Bigl[\Bigl( \frac{67}{9} -\frac{\pi^2}{3} \Bigr)\,C_A  -
   \frac{20}{9}\,T_F\, n_f \Bigr]
\,.\end{align}
For the non-cusp anomalous dimension $\gamma_x(\alpha_s)$ we only need the first coefficient, given in \eqss{H_RGE}{S_RGE}{J_RGE}.

\section{Profile Functions}
\label{app:profiles}

The optimal choice of RG scales depends on the value of $\tau$, so we use profile functions to smoothly interpolate between the small $\tau$ and large $\tau$ regions. 

Our choice of running scales is adopted from Ref.~\cite{Abbate:2010xh} with some modifications:
\begin{align} 
  \mu_H &= e_H\, Q
  \,, \\
  \mu_J(\tau) & =
  \Big[1 + e_J\, \theta(\tau_3-\tau) \Big(1 \!-\! \frac{\tau}{\tau_3}\Big)^2\,\Big]
  \sqrt{\mu_H\, \mu_\text{run}(\tau,\mu_H)}
  \,, \nn \\
  \mu_S(\tau) & = \Big[1 + e_S\, \theta(\tau_3-\tau) \Big(1 - \frac{\tau}{\tau_3}\Big)^2\,\Big] \mu_\text{run}(\tau,\mu_H)
 \,, \nn
\end{align}
where $\mu_\text{run}$ is given by
\begin{align}
& \mu_\text{run}(\tau,\mu) =
\begin{cases}
\mu_0 + a\,\tau^2/\tau_1 & \tau \leq \tau_1
\,,\\
2a\, \tau + b & \tau_1 \leq \tau \leq \tau_2
\,,\\
\mu \!-\! a (\tau\!-\!\tau_3)^2/(\tau_3 \!-\! \tau_2) & \tau_2 \leq \tau \leq \tau_3 
\,,\\
\mu & \tau > \tau_3
\,,\end{cases}
\nn \\
&
a= \frac{\mu_0-\mu}{\tau_1 -\tau_2 -\tau_3}
\,, \qquad
b = \frac{\mu \tau_1 - \mu_0 (\tau_2 + \tau_3)}{\tau_1-\tau_2-\tau_3}
\,.\end{align}
The expressions for $a$ and $b$ follow from demanding that $\mu_\mathrm{run}$ is continuous and has a continuous derivative. The value of $\mu_0$ determines the scales at $\tau = 0$, while $\tau_{1,2,3}$ determine the transition between the peak, tail, and far-tail regions discussed in \sec{resum}.  For $\tau > \tau_3$, our choice for $\mu_\mathrm{run}$ ensures that the resummation of logarithms of $\tau$ turns off.

The parameters for the central curve are
\begin{align} 
e_H =1
\,,\quad
e_B = e_S = 0
\,,\quad
\mu_0 = 2\; \text{GeV}
\,,\nn \\
\tau_1 = \frac{2\; \text{GeV}}{Q}
\,,\quad
\tau_2 = 0.15
\,,\quad
\tau_3 = 0.33
\,.\end{align}
The scale uncertainty bands are obtained by taking the envelope of the following scale variations:
\begin{align}
 a) & \quad e_H = 2^{\pm 1}\,, \quad e_J = e_S = 0
 \,,\nn \\
 b) & \quad e_H = 1\,, \quad e_J = \pm 0.5\,, \quad e_S = 0
 \,,\nn \\
 c) & \quad e_H = 1\,, \quad e_J = 0\,, \quad e_S = \pm 0.5
\,.\end{align}

\section{Plus Distributions}
\label{app:plusfunc}

The standard plus distribution for some function $g(x)$ is defined as
\begin{equation} \label{eq:plusgendef}
\bigl[\theta(x) g(x)\bigr]_+
= \lim_{\beta \to 0} \frac{\df}{\df x} \bigl[\theta(x-\beta)\, G(x) \bigr]
\end{equation}
\vspace{1ex} \ \\
with
\begin{align}
G(x) = \int_1^x\!\df x'\, g(x')
\,.\end{align}
This satisfies the boundary condition $\int_0^1 \df x\, [\theta(x) g(x)]_+ = 0$.  The two special cases we need in this paper are 
\begin{align} \label{eq:plusdef}
\cL_n(x)
&\equiv \biggl[ \frac{\theta(x) \ln^n x}{x}\biggr]_+
 \\
 &= \lim_{\beta \to 0} \biggl[
  \frac{\theta(x- \beta)\ln^n x}{x} +
  \delta(x- \beta) \, \frac{\ln^{n+1}\!\beta}{n+1} \biggr]
\,,\nn\\
\cL^\eta(x)
&\equiv \biggl[ \frac{\theta(x)}{x^{1-\eta}}\biggr]_+
 = \lim_{\beta \to 0} \biggl[
  \frac{\theta(x - \beta)}{x^{1-\eta}} +
  \delta(x- \beta) \, \frac{x^\eta - 1}{\eta} \biggr]
\,.\nn \end{align}
In our calculations, we use the plus distribution identities appearing in appendix B of Ref.~\cite{Ligeti:2008ac}. In particular, we utilize the following rescaling identity for a constant $\lambda$,
\begin{align} \label{eq:resc_plus}
  \lambda \,\cL_n(\lambda x) &= 
   \frac{\ln^{n+1}(\lambda)}{n+1} \delta(x) + \sum_{k=0}^n  \binom{n}{k}   
   \ln^{n-k}(\lambda) \cL_k(x)
   \,.
\end{align}


\bibliography{tracks}

\begin{thebibliography}{52}
\expandafter\ifx\csname natexlab\endcsname\relax\def\natexlab#1{#1}\fi
\expandafter\ifx\csname bibnamefont\endcsname\relax
  \def\bibnamefont#1{#1}\fi
\expandafter\ifx\csname bibfnamefont\endcsname\relax
  \def\bibfnamefont#1{#1}\fi
\expandafter\ifx\csname citenamefont\endcsname\relax
  \def\citenamefont#1{#1}\fi
\expandafter\ifx\csname url\endcsname\relax
  \def\url#1{\texttt{#1}}\fi
\expandafter\ifx\csname urlprefix\endcsname\relax\def\urlprefix{URL }\fi
\providecommand{\bibinfo}[2]{#2}
\providecommand{\eprint}[2][]{\url{#2}}

\bibitem[{\citenamefont{Buskulic et~al.}(1992)}]{Buskulic:1992hq}
\bibinfo{author}{\bibfnamefont{D.}~\bibnamefont{Buskulic}} \bibnamefont{et~al.}
  (\bibinfo{collaboration}{ALEPH Collaboration}), \bibinfo{journal}{Z. Phys.}
  \textbf{\bibinfo{volume}{C55}}, \bibinfo{pages}{209} (\bibinfo{year}{1992}).

\bibitem[{\citenamefont{Abreu et~al.}(1996)}]{Abreu:1996na}
\bibinfo{author}{\bibfnamefont{P.}~\bibnamefont{Abreu}} \bibnamefont{et~al.}
  (\bibinfo{collaboration}{DELPHI Collaboration}), \bibinfo{journal}{Z. Phys.}
  \textbf{\bibinfo{volume}{C73}}, \bibinfo{pages}{11} (\bibinfo{year}{1996}).

\bibitem[{\citenamefont{Dasgupta and Salam}(2004)}]{Dasgupta:2003iq}
\bibinfo{author}{\bibfnamefont{M.}~\bibnamefont{Dasgupta}} \bibnamefont{and}
  \bibinfo{author}{\bibfnamefont{G.~P.} \bibnamefont{Salam}},
  \bibinfo{journal}{J. Phys.} \textbf{\bibinfo{volume}{G30}},
  \bibinfo{pages}{R143} (\bibinfo{year}{2004}), \eprint{hep-ph/0312283}.

\bibitem[{\citenamefont{Heister et~al.}(2004)}]{Heister:2003aj}
\bibinfo{author}{\bibfnamefont{A.}~\bibnamefont{Heister}} \bibnamefont{et~al.}
  (\bibinfo{collaboration}{ALEPH Collaboration}), \bibinfo{journal}{Eur. Phys.
  J.} \textbf{\bibinfo{volume}{C35}}, \bibinfo{pages}{457}
  (\bibinfo{year}{2004}).

\bibitem[{\citenamefont{Abdallah et~al.}(2003)}]{Abdallah:2003xz}
\bibinfo{author}{\bibfnamefont{J.}~\bibnamefont{Abdallah}} \bibnamefont{et~al.}
  (\bibinfo{collaboration}{DELPHI Collaboration}), \bibinfo{journal}{Eur. Phys.
  J.} \textbf{\bibinfo{volume}{C29}}, \bibinfo{pages}{285}
  (\bibinfo{year}{2003}), \eprint{hep-ex/0307048}.

\bibitem[{\citenamefont{Achard et~al.}(2004)}]{Achard:2004sv}
\bibinfo{author}{\bibfnamefont{P.}~\bibnamefont{Achard}} \bibnamefont{et~al.}
  (\bibinfo{collaboration}{L3 Collaboration}), \bibinfo{journal}{Phys.Rept.}
  \textbf{\bibinfo{volume}{399}}, \bibinfo{pages}{71} (\bibinfo{year}{2004}),
  \eprint{hep-ex/0406049}.

\bibitem[{\citenamefont{Abbiendi et~al.}(2005)}]{Abbiendi:2004qz}
\bibinfo{author}{\bibfnamefont{G.}~\bibnamefont{Abbiendi}} \bibnamefont{et~al.}
  (\bibinfo{collaboration}{OPAL Collaboration}), \bibinfo{journal}{Eur. Phys.
  J.} \textbf{\bibinfo{volume}{C40}}, \bibinfo{pages}{287}
  (\bibinfo{year}{2005}), \eprint{hep-ex/0503051}.

\bibitem[{\citenamefont{Krohn et~al.}(2013)\citenamefont{Krohn, Lin, Schwartz,
  and Waalewijn}}]{Krohn:2012fg}
\bibinfo{author}{\bibfnamefont{D.}~\bibnamefont{Krohn}},
  \bibinfo{author}{\bibfnamefont{T.}~\bibnamefont{Lin}},
  \bibinfo{author}{\bibfnamefont{M.~D.} \bibnamefont{Schwartz}},
  \bibnamefont{and} \bibinfo{author}{\bibfnamefont{W.~J.}
  \bibnamefont{Waalewijn}}, \bibinfo{journal}{Phys. Rev. Lett.}
  \textbf{\bibinfo{volume}{110}}, \bibinfo{pages}{212001}
  (\bibinfo{year}{2013}), \eprint{arXiv:1209.2421}.

\bibitem[{\citenamefont{Waalewijn}(2012)}]{Waalewijn:2012sv}
\bibinfo{author}{\bibfnamefont{W.~J.} \bibnamefont{Waalewijn}},
  \bibinfo{journal}{Phys. Rev.} \textbf{\bibinfo{volume}{D86}},
  \bibinfo{pages}{094030} (\bibinfo{year}{2012}), \eprint{arXiv:1209.3019}.

\bibitem[{\citenamefont{Chang et~al.}(2013)\citenamefont{Chang, Procura,
  Thaler, and Waalewijn}}]{Chang:2013rca}
\bibinfo{author}{\bibfnamefont{H.-M.} \bibnamefont{Chang}},
  \bibinfo{author}{\bibfnamefont{M.}~\bibnamefont{Procura}},
  \bibinfo{author}{\bibfnamefont{J.}~\bibnamefont{Thaler}}, \bibnamefont{and}
  \bibinfo{author}{\bibfnamefont{W.~J.} \bibnamefont{Waalewijn}}
  (\bibinfo{year}{2013}), \eprint{arXiv:1303.6637}.

\bibitem[{\citenamefont{Bauer et~al.}(2000)\citenamefont{Bauer, Fleming, and
  Luke}}]{Bauer:2000ew}
\bibinfo{author}{\bibfnamefont{C.~W.} \bibnamefont{Bauer}},
  \bibinfo{author}{\bibfnamefont{S.}~\bibnamefont{Fleming}}, \bibnamefont{and}
  \bibinfo{author}{\bibfnamefont{M.~E.} \bibnamefont{Luke}},
  \bibinfo{journal}{Phys. Rev. D} \textbf{\bibinfo{volume}{63}},
  \bibinfo{pages}{014006} (\bibinfo{year}{2000}), \eprint{hep-ph/0005275}.

\bibitem[{\citenamefont{Bauer et~al.}(2001)\citenamefont{Bauer, Fleming,
  Pirjol, and Stewart}}]{Bauer:2000yr}
\bibinfo{author}{\bibfnamefont{C.~W.} \bibnamefont{Bauer}},
  \bibinfo{author}{\bibfnamefont{S.}~\bibnamefont{Fleming}},
  \bibinfo{author}{\bibfnamefont{D.}~\bibnamefont{Pirjol}}, \bibnamefont{and}
  \bibinfo{author}{\bibfnamefont{I.~W.} \bibnamefont{Stewart}},
  \bibinfo{journal}{Phys. Rev. D} \textbf{\bibinfo{volume}{63}},
  \bibinfo{pages}{114020} (\bibinfo{year}{2001}), \eprint{hep-ph/0011336}.

\bibitem[{\citenamefont{Bauer and Stewart}(2001)}]{Bauer:2001ct}
\bibinfo{author}{\bibfnamefont{C.~W.} \bibnamefont{Bauer}} \bibnamefont{and}
  \bibinfo{author}{\bibfnamefont{I.~W.} \bibnamefont{Stewart}},
  \bibinfo{journal}{Phys. Lett. B} \textbf{\bibinfo{volume}{516}},
  \bibinfo{pages}{134} (\bibinfo{year}{2001}), \eprint{hep-ph/0107001}.

\bibitem[{\citenamefont{Bauer et~al.}(2002)\citenamefont{Bauer, Pirjol, and
  Stewart}}]{Bauer:2001yt}
\bibinfo{author}{\bibfnamefont{C.~W.} \bibnamefont{Bauer}},
  \bibinfo{author}{\bibfnamefont{D.}~\bibnamefont{Pirjol}}, \bibnamefont{and}
  \bibinfo{author}{\bibfnamefont{I.~W.} \bibnamefont{Stewart}},
  \bibinfo{journal}{Phys. Rev. D} \textbf{\bibinfo{volume}{65}},
  \bibinfo{pages}{054022} (\bibinfo{year}{2002}), \eprint{hep-ph/0109045}.

\bibitem[{\citenamefont{Sj{\"o}strand et~al.}(2006)\citenamefont{Sj{\"o}strand,
  Mrenna, and Skands}}]{Sjostrand:2006za}
\bibinfo{author}{\bibfnamefont{T.}~\bibnamefont{Sj{\"o}strand}},
  \bibinfo{author}{\bibfnamefont{S.}~\bibnamefont{Mrenna}}, \bibnamefont{and}
  \bibinfo{author}{\bibfnamefont{P.}~\bibnamefont{Skands}},
  \bibinfo{journal}{JHEP} \textbf{\bibinfo{volume}{05}}, \bibinfo{pages}{026}
  (\bibinfo{year}{2006}), \eprint{hep-ph/0603175}.

\bibitem[{\citenamefont{Sj{\"o}strand et~al.}(2008)\citenamefont{Sj{\"o}strand,
  Mrenna, and Skands}}]{Sjostrand:2007gs}
\bibinfo{author}{\bibfnamefont{T.}~\bibnamefont{Sj{\"o}strand}},
  \bibinfo{author}{\bibfnamefont{S.}~\bibnamefont{Mrenna}}, \bibnamefont{and}
  \bibinfo{author}{\bibfnamefont{P.}~\bibnamefont{Skands}},
  \bibinfo{journal}{Comput. Phys. Commun.} \textbf{\bibinfo{volume}{178}},
  \bibinfo{pages}{852} (\bibinfo{year}{2008}), \eprint{arXiv:0710.3820}.

\bibitem[{\citenamefont{Farhi}(1977)}]{Farhi:1977sg}
\bibinfo{author}{\bibfnamefont{E.}~\bibnamefont{Farhi}},
  \bibinfo{journal}{Phys. Rev. Lett.} \textbf{\bibinfo{volume}{39}},
  \bibinfo{pages}{1587} (\bibinfo{year}{1977}).

\bibitem[{\citenamefont{Salam and Wicke}(2001)}]{Salam:2001bd}
\bibinfo{author}{\bibfnamefont{G.}~\bibnamefont{Salam}} \bibnamefont{and}
  \bibinfo{author}{\bibfnamefont{D.}~\bibnamefont{Wicke}},
  \bibinfo{journal}{JHEP} \textbf{\bibinfo{volume}{0105}}, \bibinfo{pages}{061}
  (\bibinfo{year}{2001}), \eprint{hep-ph/0102343}.

\bibitem[{\citenamefont{Mateu et~al.}(2013)\citenamefont{Mateu, Stewart, and
  Thaler}}]{Mateu:2012nk}
\bibinfo{author}{\bibfnamefont{V.}~\bibnamefont{Mateu}},
  \bibinfo{author}{\bibfnamefont{I.~W.} \bibnamefont{Stewart}},
  \bibnamefont{and} \bibinfo{author}{\bibfnamefont{J.}~\bibnamefont{Thaler}},
  \bibinfo{journal}{Phys. Rev.} \textbf{\bibinfo{volume}{D87}},
  \bibinfo{pages}{014025} (\bibinfo{year}{2013}), \eprint{arXiv:1209.3781}.

\bibitem[{\citenamefont{Kinoshita}(1962)}]{Kinoshita:1962ur}
\bibinfo{author}{\bibfnamefont{T.}~\bibnamefont{Kinoshita}},
  \bibinfo{journal}{J. Math. Phys.} \textbf{\bibinfo{volume}{3}},
  \bibinfo{pages}{650} (\bibinfo{year}{1962}).

\bibitem[{\citenamefont{Lee and Nauenberg}(1964)}]{Lee:1964is}
\bibinfo{author}{\bibfnamefont{T.}~\bibnamefont{Lee}} \bibnamefont{and}
  \bibinfo{author}{\bibfnamefont{M.}~\bibnamefont{Nauenberg}},
  \bibinfo{journal}{Phys. Rev.} \textbf{\bibinfo{volume}{133}},
  \bibinfo{pages}{B1549} (\bibinfo{year}{1964}).

\bibitem[{\citenamefont{Berends and Giele}(1988)}]{Berends:1987me}
\bibinfo{author}{\bibfnamefont{F.~A.} \bibnamefont{Berends}} \bibnamefont{and}
  \bibinfo{author}{\bibfnamefont{W.}~\bibnamefont{Giele}},
  \bibinfo{journal}{Nucl. Phys.} \textbf{\bibinfo{volume}{B306}},
  \bibinfo{pages}{759} (\bibinfo{year}{1988}).

\bibitem[{\citenamefont{Mangano and Parke}(1991)}]{Mangano:1990by}
\bibinfo{author}{\bibfnamefont{M.~L.} \bibnamefont{Mangano}} \bibnamefont{and}
  \bibinfo{author}{\bibfnamefont{S.~J.} \bibnamefont{Parke}},
  \bibinfo{journal}{Phys. Rept.} \textbf{\bibinfo{volume}{200}},
  \bibinfo{pages}{301} (\bibinfo{year}{1991}), \eprint{hep-th/0509223}.

\bibitem[{\citenamefont{Kosower}(1999)}]{Kosower:1999xi}
\bibinfo{author}{\bibfnamefont{D.~A.} \bibnamefont{Kosower}},
  \bibinfo{journal}{Nucl. Phys.} \textbf{\bibinfo{volume}{B552}},
  \bibinfo{pages}{319} (\bibinfo{year}{1999}), \eprint{hep-ph/9901201}.

\bibitem[{\citenamefont{Collins and Soper}(1981)}]{Collins:1981uk}
\bibinfo{author}{\bibfnamefont{J.~C.} \bibnamefont{Collins}} \bibnamefont{and}
  \bibinfo{author}{\bibfnamefont{D.~E.} \bibnamefont{Soper}},
  \bibinfo{journal}{Nucl. Phys.} \textbf{\bibinfo{volume}{B193}},
  \bibinfo{pages}{381} (\bibinfo{year}{1981}).

\bibitem[{\citenamefont{Collins and Soper}(1982)}]{Collins:1981uw}
\bibinfo{author}{\bibfnamefont{J.~C.} \bibnamefont{Collins}} \bibnamefont{and}
  \bibinfo{author}{\bibfnamefont{D.~E.} \bibnamefont{Soper}},
  \bibinfo{journal}{Nucl. Phys.} \textbf{\bibinfo{volume}{B194}},
  \bibinfo{pages}{445} (\bibinfo{year}{1982}).

\bibitem[{\citenamefont{Jain et~al.}(2011)\citenamefont{Jain, Procura, and
  Waalewijn}}]{Jain:2011xz}
\bibinfo{author}{\bibfnamefont{A.}~\bibnamefont{Jain}},
  \bibinfo{author}{\bibfnamefont{M.}~\bibnamefont{Procura}}, \bibnamefont{and}
  \bibinfo{author}{\bibfnamefont{W.~J.} \bibnamefont{Waalewijn}},
  \bibinfo{journal}{JHEP} \textbf{\bibinfo{volume}{1105}}, \bibinfo{pages}{035}
  (\bibinfo{year}{2011}), \eprint{arXiv:1101.4953}.

\bibitem[{\citenamefont{Altarelli and Parisi}(1977)}]{Altarelli:1977zs}
\bibinfo{author}{\bibfnamefont{G.}~\bibnamefont{Altarelli}} \bibnamefont{and}
  \bibinfo{author}{\bibfnamefont{G.}~\bibnamefont{Parisi}},
  \bibinfo{journal}{Nucl. Phys.} \textbf{\bibinfo{volume}{B126}},
  \bibinfo{pages}{298} (\bibinfo{year}{1977}).

\bibitem[{\citenamefont{Abbate et~al.}(2011)\citenamefont{Abbate, Fickinger,
  Hoang, Mateu, and Stewart}}]{Abbate:2010xh}
\bibinfo{author}{\bibfnamefont{R.}~\bibnamefont{Abbate}},
  \bibinfo{author}{\bibfnamefont{M.}~\bibnamefont{Fickinger}},
  \bibinfo{author}{\bibfnamefont{A.~H.} \bibnamefont{Hoang}},
  \bibinfo{author}{\bibfnamefont{V.}~\bibnamefont{Mateu}}, \bibnamefont{and}
  \bibinfo{author}{\bibfnamefont{I.~W.} \bibnamefont{Stewart}},
  \bibinfo{journal}{Phys. Rev.} \textbf{\bibinfo{volume}{D83}},
  \bibinfo{pages}{074021} (\bibinfo{year}{2011}), \eprint{arXiv:1006.3080}.

\bibitem[{\citenamefont{Catani et~al.}(1993)\citenamefont{Catani, Trentadue,
  Turnock, and Webber}}]{Catani:1992ua}
\bibinfo{author}{\bibfnamefont{S.}~\bibnamefont{Catani}},
  \bibinfo{author}{\bibfnamefont{L.}~\bibnamefont{Trentadue}},
  \bibinfo{author}{\bibfnamefont{G.}~\bibnamefont{Turnock}}, \bibnamefont{and}
  \bibinfo{author}{\bibfnamefont{B.~R.} \bibnamefont{Webber}},
  \bibinfo{journal}{Nucl. Phys.} \textbf{\bibinfo{volume}{B407}},
  \bibinfo{pages}{3} (\bibinfo{year}{1993}).

\bibitem[{\citenamefont{Korchemsky and Sterman}(1999)}]{Korchemsky:1999kt}
\bibinfo{author}{\bibfnamefont{G.~P.} \bibnamefont{Korchemsky}}
  \bibnamefont{and} \bibinfo{author}{\bibfnamefont{G.~F.}
  \bibnamefont{Sterman}}, \bibinfo{journal}{Nucl. Phys.}
  \textbf{\bibinfo{volume}{B555}}, \bibinfo{pages}{335} (\bibinfo{year}{1999}),
  \eprint{hep-ph/9902341}.

\bibitem[{\citenamefont{Fleming
  et~al.}(2008{\natexlab{a}})\citenamefont{Fleming, Hoang, Mantry, and
  Stewart}}]{Fleming:2007qr}
\bibinfo{author}{\bibfnamefont{S.}~\bibnamefont{Fleming}},
  \bibinfo{author}{\bibfnamefont{A.~H.} \bibnamefont{Hoang}},
  \bibinfo{author}{\bibfnamefont{S.}~\bibnamefont{Mantry}}, \bibnamefont{and}
  \bibinfo{author}{\bibfnamefont{I.~W.} \bibnamefont{Stewart}},
  \bibinfo{journal}{Phys. Rev.} \textbf{\bibinfo{volume}{D77}},
  \bibinfo{pages}{074010} (\bibinfo{year}{2008}{\natexlab{a}}),
  \eprint{hep-ph/0703207}.

\bibitem[{\citenamefont{Schwartz}(2008)}]{Schwartz:2007ib}
\bibinfo{author}{\bibfnamefont{M.~D.} \bibnamefont{Schwartz}},
  \bibinfo{journal}{Phys. Rev.} \textbf{\bibinfo{volume}{D77}},
  \bibinfo{pages}{014026} (\bibinfo{year}{2008}), \eprint{0709.2709}.

\bibitem[{\citenamefont{Manohar}(2003)}]{Manohar:2003vb}
\bibinfo{author}{\bibfnamefont{A.~V.} \bibnamefont{Manohar}},
  \bibinfo{journal}{Phys. Rev. D} \textbf{\bibinfo{volume}{68}},
  \bibinfo{pages}{114019} (\bibinfo{year}{2003}), \eprint{hep-ph/0309176}.

\bibitem[{\citenamefont{Bauer et~al.}(2004)\citenamefont{Bauer, Lee, Manohar,
  and Wise}}]{Bauer:2003di}
\bibinfo{author}{\bibfnamefont{C.~W.} \bibnamefont{Bauer}},
  \bibinfo{author}{\bibfnamefont{C.}~\bibnamefont{Lee}},
  \bibinfo{author}{\bibfnamefont{A.~V.} \bibnamefont{Manohar}},
  \bibnamefont{and} \bibinfo{author}{\bibfnamefont{M.~B.} \bibnamefont{Wise}},
  \bibinfo{journal}{Phys. Rev. D} \textbf{\bibinfo{volume}{70}},
  \bibinfo{pages}{034014} (\bibinfo{year}{2004}), \eprint{hep-ph/0309278}.

\bibitem[{\citenamefont{Fleming
  et~al.}(2008{\natexlab{b}})\citenamefont{Fleming, Hoang, Mantry, and
  Stewart}}]{Fleming:2007xt}
\bibinfo{author}{\bibfnamefont{S.}~\bibnamefont{Fleming}},
  \bibinfo{author}{\bibfnamefont{A.~H.} \bibnamefont{Hoang}},
  \bibinfo{author}{\bibfnamefont{S.}~\bibnamefont{Mantry}}, \bibnamefont{and}
  \bibinfo{author}{\bibfnamefont{I.~W.} \bibnamefont{Stewart}},
  \bibinfo{journal}{Phys. Rev. D} \textbf{\bibinfo{volume}{77}},
  \bibinfo{pages}{114003} (\bibinfo{year}{2008}{\natexlab{b}}),
  \eprint{arXiv:0711.2079}.

\bibitem[{\citenamefont{Manohar and Wise}(1995)}]{Manohar:1994kq}
\bibinfo{author}{\bibfnamefont{A.~V.} \bibnamefont{Manohar}} \bibnamefont{and}
  \bibinfo{author}{\bibfnamefont{M.~B.} \bibnamefont{Wise}},
  \bibinfo{journal}{Phys. Lett.} \textbf{\bibinfo{volume}{B344}},
  \bibinfo{pages}{407} (\bibinfo{year}{1995}), \eprint{hep-ph/9406392}.

\bibitem[{\citenamefont{Webber}(1994)}]{Webber:1994cp}
\bibinfo{author}{\bibfnamefont{B.}~\bibnamefont{Webber}},
  \bibinfo{journal}{Phys. Lett.} \textbf{\bibinfo{volume}{B339}},
  \bibinfo{pages}{148} (\bibinfo{year}{1994}), \eprint{hep-ph/9408222}.

\bibitem[{\citenamefont{Korchemsky and Sterman}(1995)}]{Korchemsky:1994is}
\bibinfo{author}{\bibfnamefont{G.~P.} \bibnamefont{Korchemsky}}
  \bibnamefont{and} \bibinfo{author}{\bibfnamefont{G.~F.}
  \bibnamefont{Sterman}}, \bibinfo{journal}{Nucl. Phys.}
  \textbf{\bibinfo{volume}{B437}}, \bibinfo{pages}{415} (\bibinfo{year}{1995}),
  \eprint{hep-ph/9411211}.

\bibitem[{\citenamefont{Dokshitzer and Webber}(1995)}]{Dokshitzer:1995zt}
\bibinfo{author}{\bibfnamefont{Y.~L.} \bibnamefont{Dokshitzer}}
  \bibnamefont{and} \bibinfo{author}{\bibfnamefont{B.}~\bibnamefont{Webber}},
  \bibinfo{journal}{Phys. Lett.} \textbf{\bibinfo{volume}{B352}},
  \bibinfo{pages}{451} (\bibinfo{year}{1995}), \eprint{hep-ph/9504219}.

\bibitem[{\citenamefont{Korchemsky and Tafat}(2000)}]{Korchemsky:2000kp}
\bibinfo{author}{\bibfnamefont{G.}~\bibnamefont{Korchemsky}} \bibnamefont{and}
  \bibinfo{author}{\bibfnamefont{S.}~\bibnamefont{Tafat}},
  \bibinfo{journal}{JHEP} \textbf{\bibinfo{volume}{0010}}, \bibinfo{pages}{010}
  (\bibinfo{year}{2000}), \eprint{hep-ph/0007005}.

\bibitem[{\citenamefont{Hoang and Stewart}(2008)}]{Hoang:2007vb}
\bibinfo{author}{\bibfnamefont{A.~H.} \bibnamefont{Hoang}} \bibnamefont{and}
  \bibinfo{author}{\bibfnamefont{I.~W.} \bibnamefont{Stewart}},
  \bibinfo{journal}{Phys. Lett.} \textbf{\bibinfo{volume}{B660}},
  \bibinfo{pages}{483} (\bibinfo{year}{2008}), \eprint{0709.3519}.

\bibitem[{\citenamefont{Ligeti et~al.}(2008)\citenamefont{Ligeti, Stewart, and
  Tackmann}}]{Ligeti:2008ac}
\bibinfo{author}{\bibfnamefont{Z.}~\bibnamefont{Ligeti}},
  \bibinfo{author}{\bibfnamefont{I.~W.} \bibnamefont{Stewart}},
  \bibnamefont{and} \bibinfo{author}{\bibfnamefont{F.~J.}
  \bibnamefont{Tackmann}}, \bibinfo{journal}{Phys. Rev.}
  \textbf{\bibinfo{volume}{D78}}, \bibinfo{pages}{114014}
  (\bibinfo{year}{2008}), \eprint{0807.1926}.

\bibitem[{\citenamefont{Akhoury and Zakharov}(1995)}]{Akhoury:1995sp}
\bibinfo{author}{\bibfnamefont{R.}~\bibnamefont{Akhoury}} \bibnamefont{and}
  \bibinfo{author}{\bibfnamefont{V.~I.} \bibnamefont{Zakharov}},
  \bibinfo{journal}{Phys. Lett.} \textbf{\bibinfo{volume}{B357}},
  \bibinfo{pages}{646} (\bibinfo{year}{1995}), \eprint{hep-ph/9504248}.

\bibitem[{\citenamefont{Lee and Sterman}(2007)}]{Lee:2006nr}
\bibinfo{author}{\bibfnamefont{C.}~\bibnamefont{Lee}} \bibnamefont{and}
  \bibinfo{author}{\bibfnamefont{G.~F.} \bibnamefont{Sterman}},
  \bibinfo{journal}{Phys. Rev.} \textbf{\bibinfo{volume}{D75}},
  \bibinfo{pages}{014022} (\bibinfo{year}{2007}), \eprint{hep-ph/0611061}.

\bibitem[{\citenamefont{Liu}(2011)}]{Liu:2010ng}
\bibinfo{author}{\bibfnamefont{X.}~\bibnamefont{Liu}}, \bibinfo{journal}{Phys.
  Lett.} \textbf{\bibinfo{volume}{B699}}, \bibinfo{pages}{87}
  (\bibinfo{year}{2011}), \eprint{1011.3872}.

\bibitem[{\citenamefont{Procura and Stewart}(2010)}]{Procura:2009vm}
\bibinfo{author}{\bibfnamefont{M.}~\bibnamefont{Procura}} \bibnamefont{and}
  \bibinfo{author}{\bibfnamefont{I.~W.} \bibnamefont{Stewart}},
  \bibinfo{journal}{Phys. Rev.} \textbf{\bibinfo{volume}{D81}},
  \bibinfo{pages}{074009} (\bibinfo{year}{2010}), \eprint{arXiv:0911.4980}.

\bibitem[{\citenamefont{Hahn}(2005)}]{Hahn:2004fe}
\bibinfo{author}{\bibfnamefont{T.}~\bibnamefont{Hahn}},
  \bibinfo{journal}{Comput. Phys. Commun.} \textbf{\bibinfo{volume}{168}},
  \bibinfo{pages}{78} (\bibinfo{year}{2005}), \eprint{hep-ph/0404043}.

\bibitem[{\citenamefont{Becher and Schwartz}(2008)}]{Becher:2008cf}
\bibinfo{author}{\bibfnamefont{T.}~\bibnamefont{Becher}} \bibnamefont{and}
  \bibinfo{author}{\bibfnamefont{M.~D.} \bibnamefont{Schwartz}},
  \bibinfo{journal}{JHEP} \textbf{\bibinfo{volume}{0807}}, \bibinfo{pages}{034}
  (\bibinfo{year}{2008}), \eprint{0803.0342}.

\bibitem[{\citenamefont{Thaler and Van~Tilburg}(2011)}]{Thaler:2010tr}
\bibinfo{author}{\bibfnamefont{J.}~\bibnamefont{Thaler}} \bibnamefont{and}
  \bibinfo{author}{\bibfnamefont{K.}~\bibnamefont{Van~Tilburg}},
  \bibinfo{journal}{JHEP} \textbf{\bibinfo{volume}{1103}}, \bibinfo{pages}{015}
  (\bibinfo{year}{2011}), \eprint{arXiv:1011.2268}.

\bibitem[{\citenamefont{Thaler and Van~Tilburg}(2012)}]{Thaler:2011gf}
\bibinfo{author}{\bibfnamefont{J.}~\bibnamefont{Thaler}} \bibnamefont{and}
  \bibinfo{author}{\bibfnamefont{K.}~\bibnamefont{Van~Tilburg}},
  \bibinfo{journal}{JHEP} \textbf{\bibinfo{volume}{1202}}, \bibinfo{pages}{093}
  (\bibinfo{year}{2012}), \eprint{arXiv:1108.2701}.

\bibitem[{\citenamefont{Larkoski et~al.}(2013)\citenamefont{Larkoski, Salam,
  and Thaler}}]{Larkoski:2013eya}
\bibinfo{author}{\bibfnamefont{A.~J.} \bibnamefont{Larkoski}},
  \bibinfo{author}{\bibfnamefont{G.~P.} \bibnamefont{Salam}}, \bibnamefont{and}
  \bibinfo{author}{\bibfnamefont{J.}~\bibnamefont{Thaler}}
  (\bibinfo{year}{2013}), \eprint{arXiv:1305.0007}.

\end{thebibliography}

\end{document}